\documentclass[aps,prx,twocolumn,english,superscriptaddress,amsmath,amssymb,floatfix,longbibliography]{revtex4-2}

\usepackage{graphicx} 
\usepackage{dcolumn}  
\usepackage{bm}       
\usepackage{physics}  
\usepackage{booktabs} 
\usepackage{hyperref} 
\usepackage[utf8]{inputenc}
\usepackage[T1]{fontenc}
\usepackage{ulem}

\hypersetup{
    colorlinks=true,
    linkcolor=blue,
    filecolor=magenta,      
    urlcolor=blue,
    citecolor=blue,
}

\begin{document}

\title{Transition from Statistical to Hardware-Limited Scaling in Photonic Quantum State Reconstruction}

\author{Attila Baumann}
\affiliation{Institute for Physics and Astronomy, Eötvös Loránd University, Pázmány Péter s 1/A, Budapest 1117, Hungary}

\author{Zsolt Kis}
\affiliation{Institute for Solid State Physics and Optics, HUN-REN Wigner Research Centre for Physics, Konkoly-Thege Miklós út 29-33, Budapest 1121, Hungary}

\author{János Koltai}
\affiliation{Institute for Physics and Astronomy, Eötvös Loránd University, Pázmány Péter s 1/A, Budapest 1117, Hungary}

\author{Gábor Vattay}
\email[Contact author: ]{gabor.vattay@ttk.elte.hu}
\affiliation{Institute for Physics and Astronomy, Eötvös Loránd University, Pázmány Péter s 1/A, Budapest 1117, Hungary}

\date{\today}

\begin{abstract}
The theoretical efficiency of classical shadow tomography is predicated on a perfect Haar-random unitary ensemble, yet this mathematical ideal remains physically unattainable in near-term hardware.
Here, we report the experimental characterization of a hardware-induced systematic error floor on integrated photonic processors: an operational ``Hardware Horizon'' where the reconstruction error undergoes a sharp crossover.
While the error initially obeys the predicted statistical scaling $\mathcal{O}(M^{-1/2})$ in a variance-dominated regime, it abruptly saturates at a floor determined by the spectral distortions of the realized unitary group, marking a transition to a bias-dominated regime.
By deriving a phenomenological error model, we decouple the competing mechanisms of static coherent spectral distortion and dynamic decoherence, demonstrating that this device-specific noise floor imposes a systematic saturation limit that statistical accumulation cannot overcome.
These findings establish that the utility of shadow tomography on NISQ (noisy intermediate-scale quantum) hardware is defined by a specific scaling law involving hardware parameters, necessitating active compensation strategies to bridge the gap between theoretical purity and the noisy reality of integrated photonics.
\end{abstract}

\maketitle

\section{Introduction}
\label{sec:introduction}

Efficiently characterizing quantum states is a fundamental prerequisite for scaling quantum information technologies~\cite{Huang2020}.
Standard Quantum Tomography (SQT), however, faces an insurmountable barrier for large systems, as its resource cost scales exponentially with the number of components~\cite{hradil1997quantum, thew2002qudit}.
This ``scaling problem'' has necessitated the development of resource-efficient protocols, including compressed sensing, matrix-product-state tomography, and permutationally-invariant tomography~\cite{gross2010quantum, cramer2010efficient}.
Among these, classical shadow tomography has emerged as a versatile alternative~\cite{Aaronson2018}.
By averaging measurements performed after applying a series of random unitary transformations theoretically drawn from a continuous Haar distribution this technique allows for the reconstruction of the full density matrix with a measurement count that scales only polynomially with system size~\cite{Huang2020}.

Photonic processors represent a leading platform for implementing such protocols~\cite{slussarenko2019photonic}.
Devices based on silicon-nitride waveguides offer scalability, high-speed operation, and excellent coherence due to photon isolation at room temperature~\cite{shekhar2024roadmapping, Roeloffzen2018}.
With appropriate design, these processors can implement complex unitary transformations by propagating photons through a tunable mesh of phase shifters and beam splitters, effectively realizing the random rotations required for shadow tomography.

However, a critical disparity exists between the mathematical axioms of shadow tomography and the physical reality of NISQ implementations.
The protocol’s polynomial scaling advantage is predicated on sampling from a perfect Haar-random ensemble---an assumption that is physically unattainable in near-term hardware.
Real-world photonic processors do not implement continuous group operations; instead, they realize a spectrally distorted approximation of the unitary group, governed by the inevitable interplay of static coherent spectral distortion and dynamic thermal noise.
This creates a structural incompatibility: while statistical errors vanish as $\mathcal{O}(M^{-1/2})$, these hardware-induced spectral distortions impose a systematic, non-vanishing floor on reconstruction fidelity.
Consequently, the central challenge shifts from optimizing sample complexity to characterizing this operational ``Hardware Horizon,'' which defines the specific resolution limit of the utilized device.

In this paper, we address this gap by reporting the experimental characterization of a hardware-dependent systematic error floor---an operational ``Hardware Horizon''---for shadow tomography on integrated photonics.
We implement a classical analogue of density matrix reconstruction on a programmable eight-channel QuiX Quantum photonic processor~\cite{QuixQuantum2025}.
By contrasting extensive numerical simulations with experimental data, we identify two distinct scaling regimes.
In the \textit{Statistical Regime}, reconstruction error decreases according to the theoretical prediction $\mathcal{O}(M^{-1/2})$, characteristic of a variance-dominated regime.
However, we observe a crossover into a \textit{Device Regime}, a bias-dominated regime where the fidelity saturates despite increased measurement statistics ($M$).
This saturation reveals that the reconstruction is no longer limited by sample size, but by the intrinsic spectral distortion of the realized unitary group.
Through a phenomenological error model, we decouple the mechanisms driving this saturation, demonstrating that it arises from the competition between static coherent spectral distortion and dynamic decoherence.
We establish that this device-specific error floor imposes a systematic saturation limit on accuracy that statistical accumulation cannot overcome.
These findings suggest that the utility of shadow tomography on NISQ hardware is defined by a specific scaling law involving hardware parameters, necessitating the development of active compensation strategies to bridge the gap between the theoretical purity of the Haar measure and the noisy reality of integrated photonics.

The remainder of this paper is structured as follows: In Section~\ref{sec:theory}, we describe our methodology, including the experimental setup on the photonic chip and the core principles of the reconstruction algorithm.
Section~\ref{sec:numerics} presents the numerical simulations that validate the analytical model and establish the theoretical baseline.
In Section~\ref{sec:expres}, we detail the experimental results, deriving the phenomenological error model that defines the Hardware Horizon.
Finally, in Section~\ref{sec:conclusion}, we discuss the implications of these hardware-imposed limits for future quantum state characterization.
The detailed theoretical derivations for the reconstruction formula, Haar-measure generation, and the analytical Frobenius norm are provided in Appendices  \ref{subsec:reconstruction_theory}, \ref{app:haar_measure} and \ref{app:frobenius}.

\section{Methods}\label{sec:theory}

The central concept of this work involves reconstructing the quantum density matrix $\rho$ solely from classical measurement data.
We consider a $d$-dimensional quantum system where, prior to detection, a unitary transformation $U$ is applied, effectively rotating the measurement basis relative to the state.
The measurement outcome is formalized through the binary indicator variable $b_i$.
Specifically, the value $b_i=1$ corresponds to the event where the system collapses into the $i$-th state of the computational basis, while $b_j=0$ for all $j \neq i$.
The probability $P(b_i=1)$ represents the likelihood of this specific event occurring.
This probability is determined by the Born rule, which dictates that the likelihood of observing outcome $i$ is equal to the $i$-th diagonal element of the density matrix in the rotated frame:
\begin{equation}
    P(b_i=1) = (U\rho U^{\dagger})_{ii} = \sum_{j,k} U_{ij}\rho_{jk}U_{ik}^{*}.
\end{equation}
For a fixed $U$, a set of measurements yields the probabilities $P(b_i=1)$.
These outcomes allow us to define a diagonal matrix $\rho^{\text{exp}}$, where $\rho_{ii}^{\text{exp}} = P(b_i=1)$.
Consequently, an initial estimator $\hat{\rho}^{(1)}$ of the original density matrix elements can be constructed as:
\begin{equation}
    \hat{\rho}_{ab}^{(1)} = \sum_{i} U_{ia}^{*} \rho_{ii}^{\text{exp}} U_{ib} = \sum_{i,j,k} U_{ia}^{*} U_{ij} \rho_{jk} U_{ik}^{*} U_{ib}.
\end{equation}
To generalize this, we repeat the procedure with random matrices drawn from the Haar distribution of the compact group of unitary matrices ${\cal U}(d)$.
Averaging over this ensemble yields:
\begin{equation}
    I_{ab} = \mathbb{E}_{U}[\hat{\rho}_{ab}^{(1)}] = \int_{U(d)} dU \sum_{i,j,k} U_{ia}^{*} U_{ij} \rho_{jk} U_{ik}^{*} U_{ib},
\end{equation}
where $dU$ denotes the Haar measure.
The evaluation of this integral requires computing fourth-order moments of the form $\int dU U_{a'b'} U_{c'd'} U_{e'f'}^{*} U_{g'h'}^{*}$, which may be performed using Weingarten calculus~\cite{Collins2022weingarten}.
This calculation results in:
\begin{equation}
    I_{ab} = \frac{1}{d+1} (\rho_{ab} + \delta_{ab}).
\end{equation}
Therefore, the elements of the original density matrix $\rho_{ab}$ can be expressed directly in terms of the averaged quantity $I_{ab}$:
\begin{equation}
    \rho_{ab} = (d+1)I_{ab} - \delta_{ab}.\label{eq:recform}
\end{equation}

In a practical experimental setting, $I_{ab}$ is estimated by averaging $\rho_{ab}^{(k)} = \sum_{i} U_{ia}^{*} \rho_{ii}^{\text{exp}} U_{ib}$ over a finite number $M$ of random unitary transformations $\{U^{(k)}\}_{k=1}^M$:
\begin{equation}
    \overline{I_{ab}}^{(M)} = \frac{1}{M}\sum_{k=1}^{M}\rho_{ab}^{(k)} = \mathbb{E}_{U}^{(M)}\left[\sum_i U^*_{ia} \rho_{ii}^{\text{exp}} U_{ib}\right].
\label{eq:rho_reconstructed_final}
\end{equation}
The final reconstructed matrix is thus given by:
\begin{equation}
    \rho_{ab}^{(\text{rec})}=(d+1)\overline{I_{ab}}^{(M)}-\delta_{ab}.
\label{eq:reconstruction}
\end{equation}
While this formula provides the reconstruction mechanism, the critical limitation lies in the statistical convergence;
specifically, how the reconstruction error scales with the dimension $d$ when the sample size $M$ remains finite.
The novelty of our approach is that, unlike the standard quantum approach, where each measurement yields a single discrete outcome $\hat{b} \in \{1, \dots, d\}$ (a ``click''), in our photonic implementation, we do not observe a single collapse.
Instead, we measure the probability of all $d$ outputs simultaneously.
This provides the full probability distribution $p_k = \langle k | U \rho U^\dagger |k \rangle$ for a given unitary $U$ in a single experimental run.
The estimator for a single snapshot is rank-1:
\begin{equation}
    \hat{\rho}_{\text{single}} = (d+1) U^\dagger |\hat{b}\rangle\langle\hat{b}| U - \mathbb{I}.
\end{equation}
This process introduces significant shot noise because information about the other $d-1$ outcomes is lost in the collapse.
The estimator becomes a weighted sum:
\begin{equation}
    \hat{\rho}_{\text{int}} = (d+1) \sum_{k=1}^d p_k U^\dagger |k\rangle\langle k| U - \mathbb{I}.
\end{equation}
By accessing the full distribution, the shot noise associated with outcome selection is eliminated, leaving only the variance from the Haar random sampling of $U$.
As derived in Appendix~\ref{app:frobenius}, the Mean Squared Error scales linearly with $d$:
\begin{equation}
    \mathbb{E}\left[ \| \hat{\rho}^{(M)} - \rho \|_F^2 \right]_{\text{int}} = \frac{d\, \text{Tr}(\rho^2)-1}{M}.
\end{equation}
In the following section, we detail the realization and experimental validation of this scheme on a photonic integrated circuit.

\subsection{Experimental realization}

\begin{figure}[htbp]
    \centering
    \includegraphics[width=0.98\linewidth]{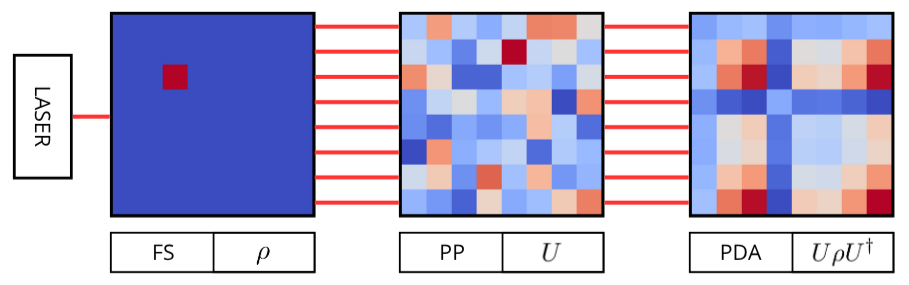}
    \caption{Schematic of the experimental arrangement.
The LASER source acts as the carrier, while the Fiber Switch (FS) prepares the initial state, denoted by the density matrix $\rho$.
This state propagates through the Photonic Processor (PP), which implements the specific unitary transformation $U$.
Finally, the Photonic Diode Array (PDA) detects the output intensity distribution, generating the input for the reconstruction of the rotated density matrix.}
    \label{fig:expsetup}
\end{figure}

The experiments were conducted using an 8-channel Mach-Zehnder interferometer (MZI) mesh, following the architecture proposed by Clements et al.~\cite{clements2016optimal} and implemented via a fully programmable Quix Quantum processor~\cite{taballione2021universal}.
The experimental setup, illustrated in Figure~\ref{fig:expsetup}, comprises a $1550\,\text{nm}$ laser source (Thorlabs S5FC1005P) and a fiber switch that governs photon injection into specific channels of the 8-channel processor.

The chosen light source is a continuous-wave Fabry-Perot laser diode operating at a central wavelength of 1550 nm with a -3dB bandwidth of 50 nm. The coherence length of this broadband field is defined by $L_c = \lambda^2 / \Delta\lambda$, yielding approximately 48 µm in vacuum. Accounting for the refractive index of the silicon nitride host material ($n \approx 2$ at 1550 nm), the effective coherence length within the waveguides is reduced to approximately 24 µm. While this corresponds to a relatively short coherence window of roughly 31 wavelengths, it does not degrade the implementation of the unitary transformations. The fundamental building block of the integrated photonic network is the balanced Mach-Zehnder interferometer, composed of tunable beam splitters and phase shifters, see Fig. 2.

This specific architecture is engineered such that the structural optical path length remains symmetric regardless of the internal routing. Because the intrinsic path-length differences between the respective interferometer arms are negligible compared to the 24 µm coherence length, the network sustains highly stable interference across all physical modes, ensuring the operational fidelity of the setup.

This macroscopic regime allows us to treat the detection voltages at the photodiode array strictly as classical probabilities, effectively isolating the statistical scaling behavior to the variance introduced by the Haar-random sampling of the unitary matrices.
Output intensities are subsequently recorded by an array of photodiodes (Thorlabs FGA01FC).
Both the photonic processor and the photodiode array are directly interfaced to a control computer.
For measurement protocols, we employ Python scripts that leverage the proprietary ``dogwood'' library from Quix Quantum;
this framework not only supports basic chip operations but also enables the precise configuration of arbitrary unitary matrices.

\begin{figure}[htbp]
    \centering
    \includegraphics[width=0.5\linewidth]{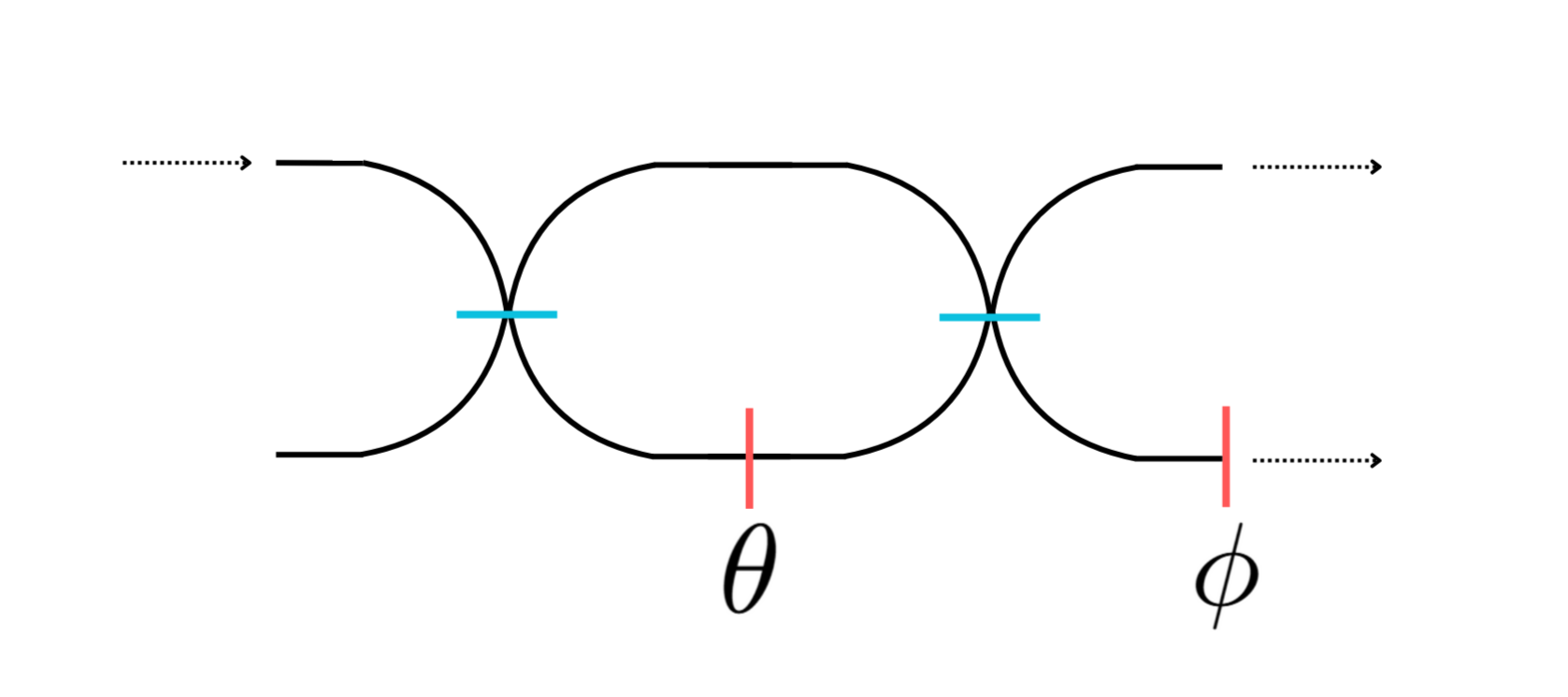}
    \caption{Sketch of the unit cell.
Phase sifters $\phi$ and $\theta$ are denoted by red lines and the 50:50 beam splitter by blue lines.}
    \label{fig:unit_cell}
\end{figure}

A unit cell of the MZI, shown in Fig.~\ref{fig:unit_cell}, consists of a phase shifter with an external phase $\phi$ and an internal interferometer phase $\theta$, together with a pair of 50:50 beam splitters.

The unit cell acts on the input channels as the unitary matrix:
$$
    S(\theta, \phi)
= \frac{1}{2}
\begin{pmatrix}
1 - e^{-i\theta} & -i(e^{-i\theta} + 1)e^{-i\phi} \\[4pt]
-i(e^{-i\theta} + 1)e^{-i\phi} & -(1 - e^{-i\theta})e^{-i\phi}
\end{pmatrix}.
$$
The photonic chip comprises a vertically arranged structure of four layers, each extending to a depth of eight unit cells.
This configuration provides the necessary degrees of freedom to implement arbitrary unitary matrices.
By modulating the waveguide phases over the full $2\pi$ range, any transformation within the ${\cal SU}(2)$ unitary group can be realized at the unit cell level (for the mathematical construction, see Ref. \cite{taballione2021universal}).
These phase shifters are actuated via the thermo-optic effect using resistive heating;
consequently, every unitary matrix maps to a distinct array of control voltages.
By composing the transfer matrices of the individual unit cells, one can derive the global transfer and scattering matrices of the full system.
This implementation achieves a fidelity of 98\%, ensuring that the realized matrices exhibit minimal distortion compared to the target transformations.
Detailed specifications of the system are provided in Ref. \cite{Taballione:19}.

\subsection{Measurement protocols}\label{sec:protocols}

A fundamental distinction of our measurement protocol compared to standard classical shadow tomography lies in the nature of the detection phase.
Conventional quantum implementations rely on single-shot projective measurements, which yield a single discrete outcome per experimental run.
Such approaches inevitably embed projective shot noise into the acquired dataset.
Our experimental setup, conversely, acquires the full probability distribution simultaneously through the continuous detection of macroscopic optical intensities across all output channels.
By extracting the exact outcome probabilities in a single measurement cycle, the protocol entirely circumvents the shot noise inherent in quantum state collapse.
The statistical variance present in our state reconstruction is thus isolated strictly to the finite Haar-random sampling of the unitary matrices.
This methodological choice permits a precise evaluation of the scaling limit, free from the obscuring effects of outcome projection.
The experimental procedure functions as a classical analogue to quantum shadow tomography.
It relies on the precise manipulation of spatial modes to encode quantum information, followed by random unitary rotations and intensity detection.
The protocol proceeds in three distinct phases:

\begin{enumerate}
    \item \textbf{State Initialization (Injection):} The input quantum state is defined by the spatial distribution of the laser light injected into the processor.
Physically, injecting coherent laser light solely into the $a$-th input waveguide prepares the computational basis state $\ket{a}$.
For example, injection into channel $a=3$ prepares the state $\ket{\psi} = (0,0,1,0,0,0,0,0)^T$, corresponding to a density matrix $\rho$ where $\rho_{33}=1$ and all other elements vanish.
    \item \textbf{Unitary Evolution:} Using the control library, we apply a specific unitary transformation, $U$, to the photonic processor.
This transformation is realized by tuning the mesh of Mach-Zehnder interferometers.
For shadow tomography, these unitaries are drawn from the Haar measure (or a specific experimental design as detailed below).
    \item \textbf{Detection and Probability Assignment:} The output optical intensities are recorded by the photodiode array.
Since the detectors measure classical intensity rather than single-photon counts in this regime, we derive the quantum probabilities via normalization.
Let $V_i$ denote the voltage measured at the $i$-th output photodiode, which is proportional to the light intensity.
The probability of observing the system in state $\ket{b_i}$ (the $i$-th output mode) is calculated as:
    \begin{equation}
        \rho^{\text{exp}}_{ii} = P(b_i) = \frac{V_i}{\sum_{j=1}^{8} V_j}.
    \end{equation}
    This normalized vector represents the diagonal of the density matrix in the rotated basis, serving as the input for the reconstruction algorithm.
\end{enumerate}

To rigorously benchmark the platform's fidelity, we executed four distinct experimental protocols.
These progress from full-dimensional trivial states to complex superpositions embedded in subspaces:

\begin{description}    
    \item[\textbf{I. The 8-Dimensional Trivial State}] 
    This protocol calibrates the system's ground truth.
We inject laser light into the first channel, preparing the fundamental state $\ket{1,0,0,0,0,0,0,0}$.
The chip is subjected to a sequence of Haar-random unitaries, $U_{\text{Haar};8\times8}$.
A successful reconstruction yields the initial projector:
    \[
        \rho_{\text{triv};8\times8}^{(\text{rec})} = \rho_{\text{triv};8\times8} = \text{diag}(1,0,0,0,0,0,0,0).
    \]

    \item[\textbf{II. The 8-Dimensional Random State}]
    To reconstruct an arbitrary coherent superposition, we introduce a fixed randomizing unitary, $U_{\text{rdm};8\times8}$.
While the input remains channel 1, the effective measurement unitary applied to the chip is the composition:
    \[
        U_{\text{meas};8\times8} = U_{\text{Haar};8\times8} \, U_{\text{rdm};8\times8}.
    \]
    Significantly, the reconstruction algorithm is provided only with the sequence $U_{\text{Haar};8\times8}$.
If successful, the algorithm reconstructs the state rotated by the hidden randomization matrix:
    \[
        \rho_{\text{rdm};8\times8}^{(\text{rec})} = U_{\text{rdm};8\times8} \, \rho_{\text{triv};8\times8} \, U_{\text{rdm};8\times8}^\dagger.
    \]

    \item[\textbf{III. The 4-Dimensional Trivial State}]
    We verify control over isolated subspaces by restricting operations to the first four modes.
We generate Haar-random unitaries $U_{\text{Haar};4\times4}$ and embed them into the full $8\times 8$ processor via the direct sum:
    \[
        U_{\text{meas};8\times8} = 
        \begin{pmatrix} 
            U_{\text{Haar};4\times4} & 0 \\ 
            0 & \mathbb{I}_{4\times4} 
        \end{pmatrix}.
    \]
    Measurements are restricted to the first four output channels, with normalization calculated over that subspace ($\sum_{i=1}^4 V_i$).
This tests the system's ability to decouple from the ``environment'' (modes 5-8).
The target reconstruction is:
    \[
        \rho_{\text{triv};4\times4}^{(\text{rec})} = \text{diag}(1,0,0,0).
    \]

    \item[\textbf{IV. The 4-Dimensional Random State}]
    Finally, we combine subspace embedding with arbitrary rotation.
We generate a random unitary $U_{\text{rdm};4\times4}$ and compose it with the Haar sampling sequence within the upper-left block:
    \[
        U_{\text{meas};8\times8} = 
        \begin{pmatrix} 
            U_{\text{Haar};4\times4}U_{\text{rdm};4\times4} & 0 \\ 
            0 & \mathbb{I}_{4\times4} 
        \end{pmatrix}.
    \]
    By recording the full composition $U_{\text{meas}}$ on the device but using only $U_{\text{Haar};4\times4}$ for reconstruction data, we recover the rotated state confined strictly to the target subspace:
    \[
        \rho_{\text{rdm};4\times4}^{(\text{rec})} = U_{\text{rdm};4\times4} \, \rho_{\text{triv};4\times4} \, U_{\text{rdm};4\times4}^\dagger.
    \]
\end{description}

\section{The statistical regime: Ideal Haar scaling}\label{sec:numerics}

Prior to deployment on the photonic hardware, we conducted extensive numerical simulations to validate the reconstruction protocols detailed in Section~\ref{sec:theory}.
These simulations emulate the classical analogue of shadow tomography by calculating the exact intensity distributions expected from the ideal unitary operations, thereby establishing a theoretical baseline free from hardware imperfections.
The primary objective was to substantiate the analytical relationship between the sample size, $M$, of applied unitary transformations and the reconstruction error, as derived in Appendix~\ref{app:frobenius}.
This step is critical for optimizing the experimental resource budget.

\subsection{Error quantification metrics}

\begin{figure}[htbp]
    \centering
    \includegraphics[width=1.1\linewidth]{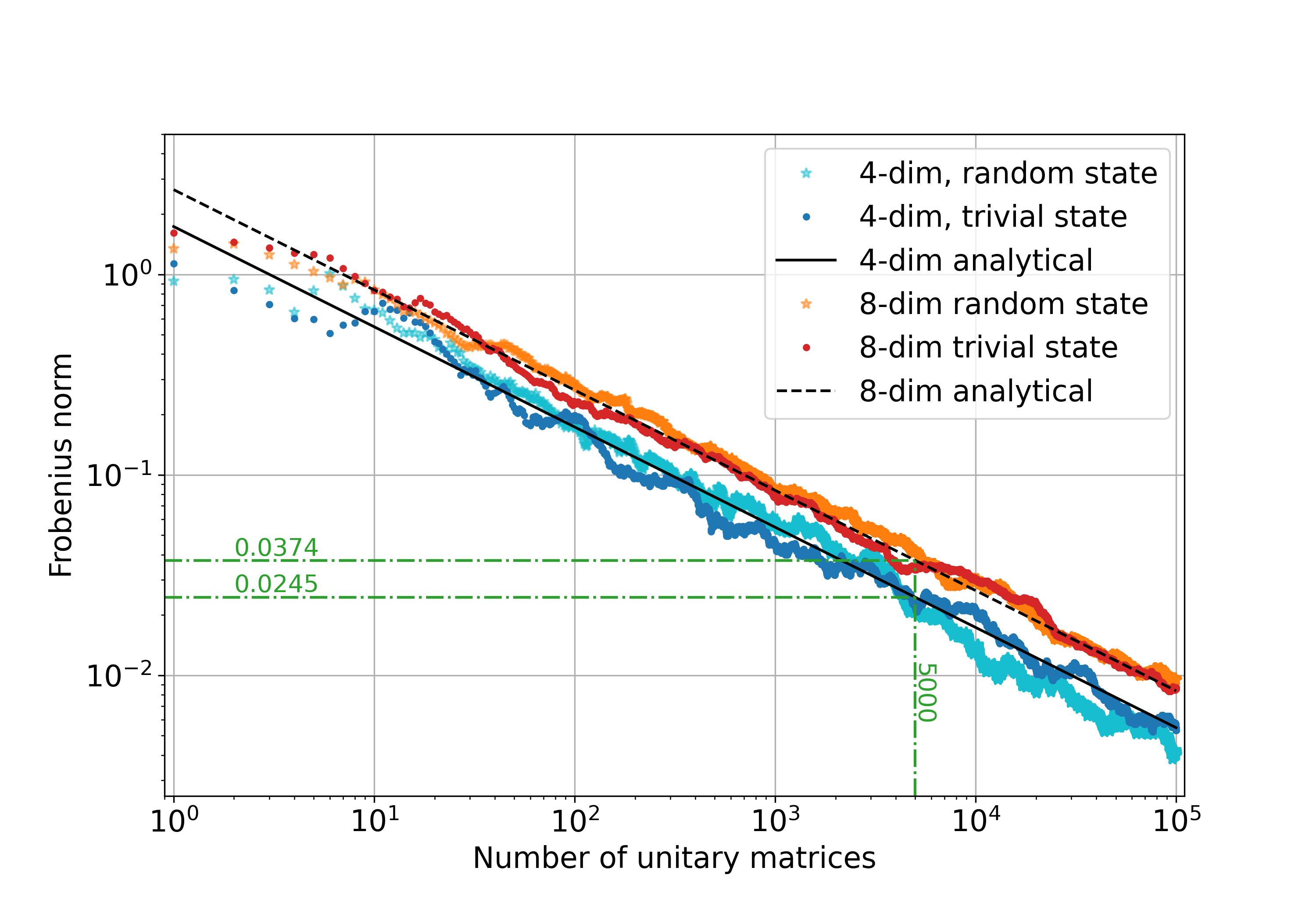}
    \caption{Scaling of reconstruction error.
The plot shows the dependence of the Frobenius norm on the number of unitary measurements ($M$) for both 4-dimensional and 8-dimensional systems.
The simulated data (points) adhere strictly to the analytical prediction $\sqrt{(d-1)/M}$ (lines).
The dash-dotted green lines mark the theoretical accuracy limit for the experimentally chosen sample size of $M=5000$.}
    \label{fig:numfig2}
\end{figure}

Quantifying the deviation between the reconstructed density matrix, $\rho^{(\text{rec})}$, and the target state, $\rho$, requires a robust metric.
While Fidelity and Trace Distance are standard measures in quantum information~\cite{jozsa1994fidelity, nielsen2000quantum}, they present specific limitations in this context.
Due to finite statistics and numerical precision, the reconstructed matrix $\rho^{(\text{rec})}$ is not guaranteed to be positive semi-definite.
Consequently, fidelity calculations can yield unphysical values exceeding unity, rendering them unreliable for precise error scaling.
Conversely, the trace distance, while a valid metric, focuses on distinguishability and may not adequately capture the accuracy of individual off-diagonal elements (coherences), which are of central interest in photonic interferometry.
Therefore, we employ the Frobenius norm of the difference matrix as our primary figure of merit.
It provides a direct, element-wise quantification of the reconstruction accuracy:
\begin{equation}
    \|\Delta \rho\|_F \equiv \|\rho^{(\text{rec})} - \rho\|_F = \sqrt{\sum_{i=1}^d \sum_{j=1}^d \left| \rho^{(\text{rec})}_{ij} - \rho_{ij} \right|^2}.\label{eq:Frobenius}
\end{equation}
As derived in Appendix~\ref{app:frobenius}, the expected value of this error scales inversely with the square root of the measurement count:
\begin{equation}
    \mathbb{E}\left[ \|\Delta \rho\|_F^2 \right] = \frac{d\, \text{Tr}(\rho^2)-1}{M}.\label{eq:Frobeniuserrgen}
\end{equation}
In the context of our measurement protocols, the original density matrix represented a pure state $\rho=\ket{\psi}\bra{\psi}$ for which $\text{Tr}(\rho^2)=1$, yielding
\begin{equation}
    \mathbb{E}\left[ \|\Delta \rho\|_F^2 \right] = \frac{d-1}{M}.\label{eq:Frobeniuserr}
\end{equation}

\begin{figure}[htbp]
    \centering
    \includegraphics[width=0.58\linewidth]{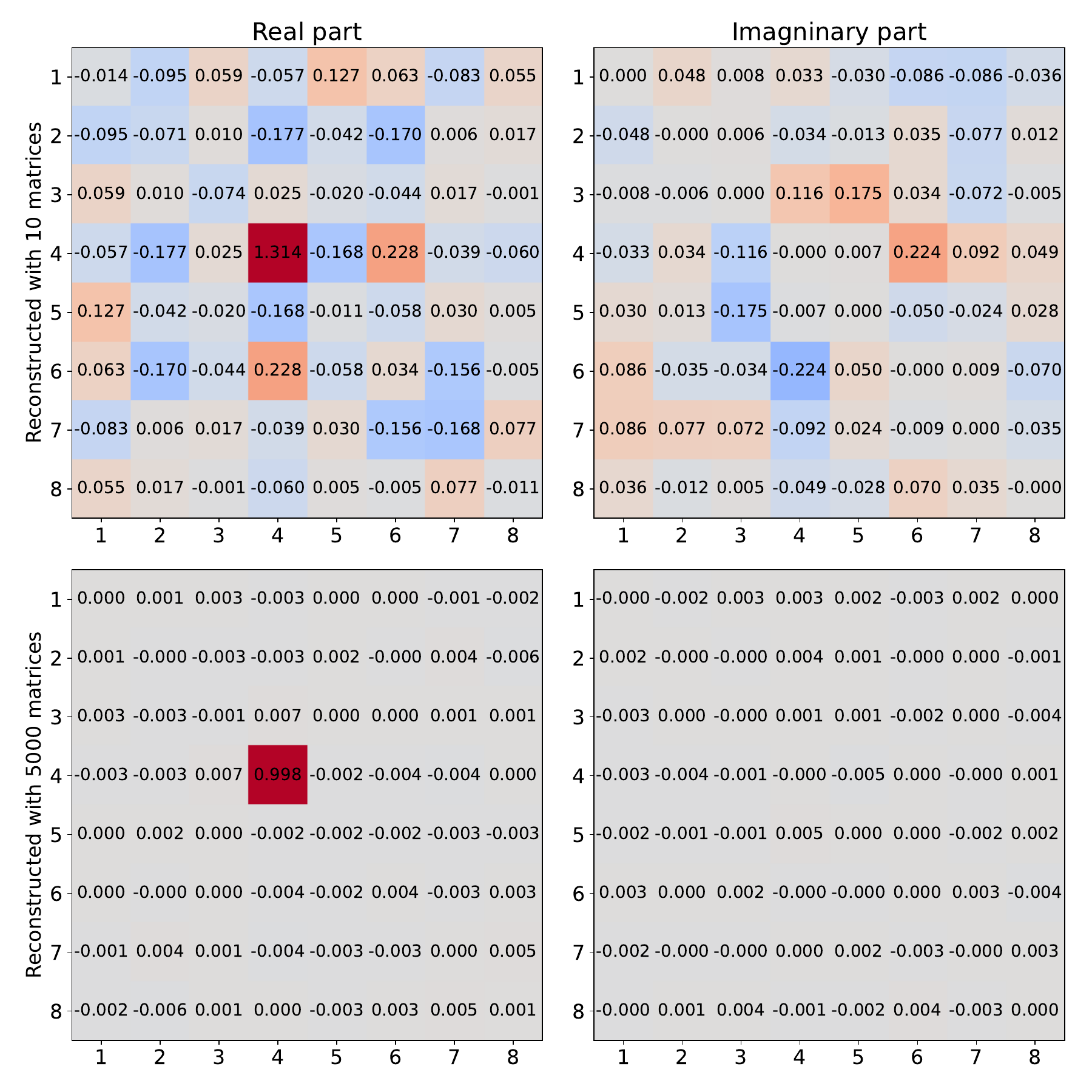}
    \hfill
    \includegraphics[width=0.39\linewidth]{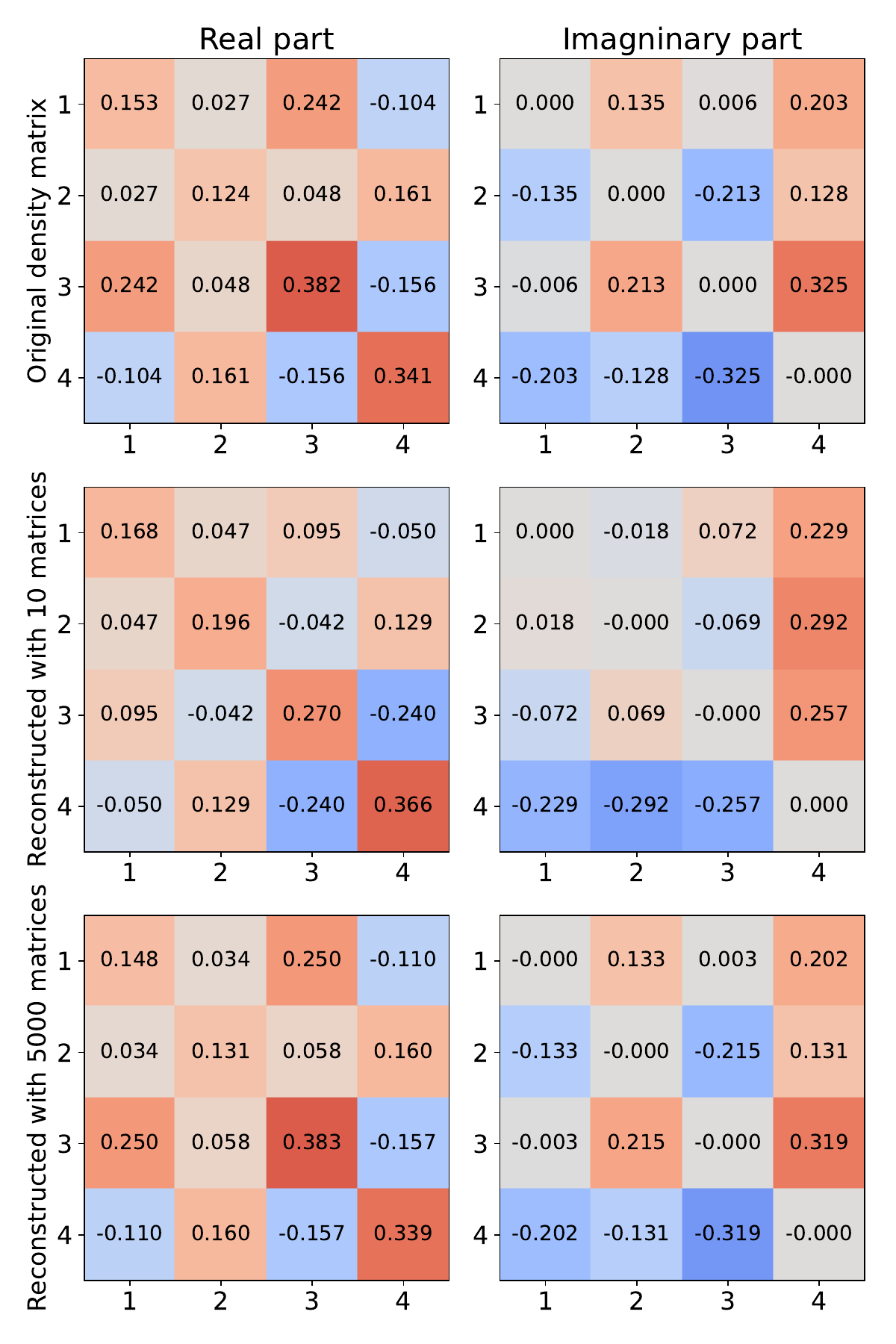}
    \caption{Numerical validation of density matrix reconstruction.
\textbf{Left:} Reconstruction of an 8-dimensional trivial state. \textbf{Right:} Reconstruction of a 4-dimensional random state.
The heatmaps display the real and imaginary components of the density matrices for the original random state ($4 \times 4$ case) and after $M=10$ and $M=5000$ (both cases) unitary transformations.
The numerical values within the cells indicate the reconstructed amplitudes, demonstrating convergence toward the target state.}
    \label{fig:numfig1ab}
\end{figure}

\subsection{Simulation results}

The simulations mirror the four experimental configurations (I--IV) established previously.
By substituting the simulated ideal intensities into the reconstruction formula (Eq.~\ref{eq:rho_reconstructed_final}), we verified the protocol for both trivial and random states in 4-dimensional and 8-dimensional Hilbert spaces.
Figure~\ref{fig:numfig1ab} visualizes the convergence of the reconstruction for an $8\times 8$ trivial state and a $4\times 4$ random state.
The heatmaps and corresponding matrix elements demonstrate that with $M=5000$ unitary samples, the reconstructed elements approach the target values with high precision.
The quantitative scaling behavior is presented in Figure~\ref{fig:numfig2}. Here, we plot the Frobenius norm against the number of unitary transformations $M$ on a logarithmic scale, extending to $M=10^5$.
The simulation data aligns perfectly with the analytical prediction (solid lines), confirming the $\mathcal{O}(M^{-1/2})$ scaling law.
These simulations define the Statistical Regime, where reconstruction fidelity is bounded solely by sample size $M$.
The specific values 0.037 (8-dim) and 0.024 (4-dim) represent the asymptotic `zero-noise' limit.
Any deviation from this $\mathcal{O}(M^{-1/2})$ scaling in the physical apparatus identifies the onset of the Device Regime.

\section{Experimental results}\label{sec:expres}
\begin{figure}[htbp]
    \centering
    \includegraphics[width=1.1\linewidth]{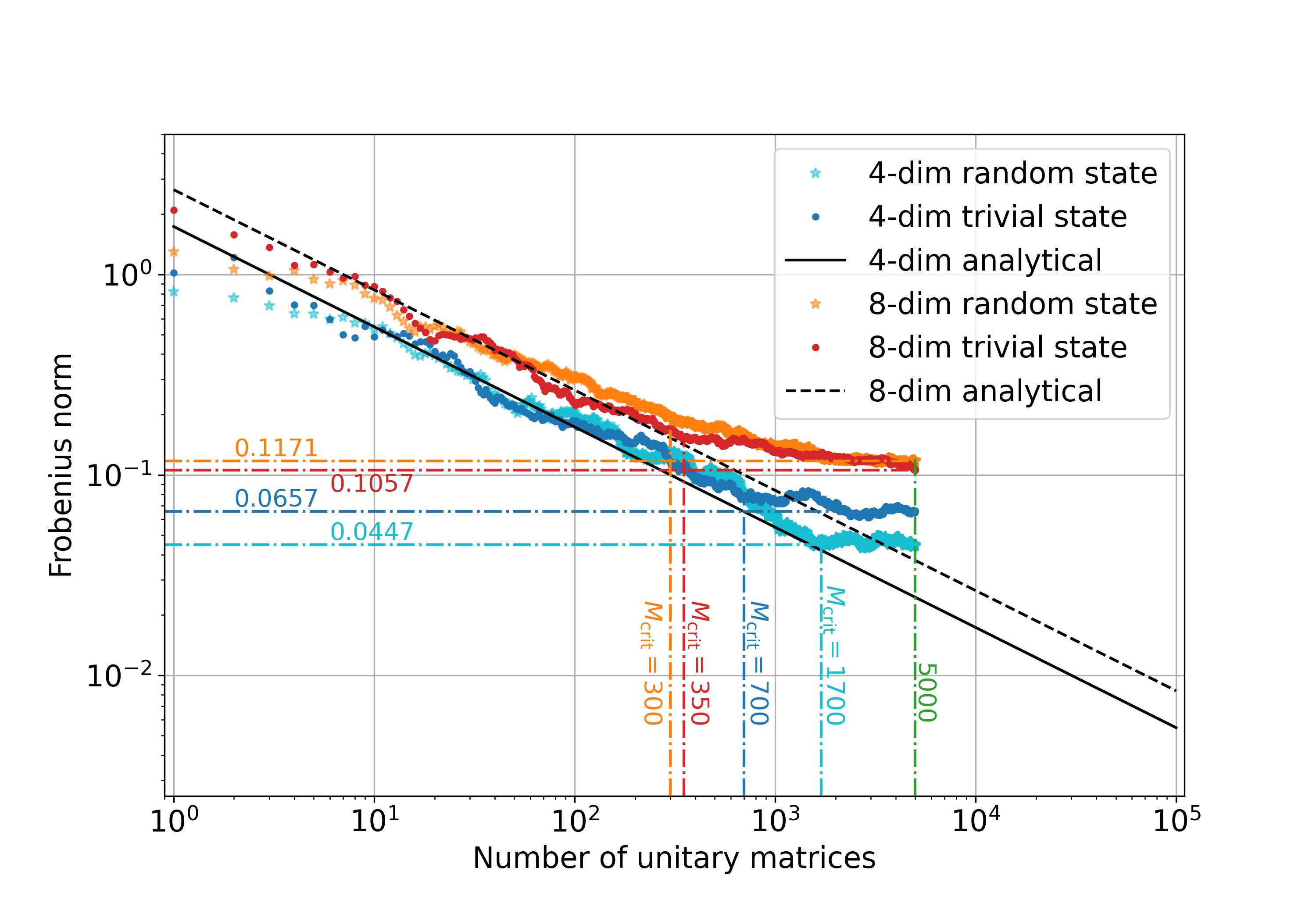}
    \caption{\textbf{Observation of the Hardware Horizon.} The scaling of reconstruction error (Frobenius norm) versus measurement count ($M$) for 4- and 8-dimensional states.
The data initially follows the predicted statistical scaling law $\mathcal{O}(M^{-1/2})$ (solid black lines).
However, a distinct crossover occurs at a critical sample size ($M_{crit}$), transitioning from a variance-dominated to a bias-dominated regime where the error diverges from the theoretical prediction and saturates at a hardware-imposed floor.
This plateau defines the ``Hardware Horizon,'' governed by the intrinsic spectral distortion of the photonic processor, which prevents convergence to the theoretical accuracy limit (green dash-dotted lines).}
    \label{fig:expfig2}
\end{figure}

Figure \ref{fig:expfig2} compares the experimental data with the analytical curves.
A clear divergence between the expected and experimental Frobenius norms becomes apparent when averaging 5000 unitary matrices.
Specifically, the experimental Frobenius norm saturates to a finite value, whereas the theoretical curve continues to decrease.
Consequently, the reconstructed density matrix no longer converges to the input density matrix.
This deviation is not attributable to statistical errors arising from the finite number of measurement samples.
Rather, the discrepancy stems from two primary sources. The first source relates to static imperfections in the chip.
Implementing unitary matrices on the processor necessitates the fine-tuning of beam splitters and phase shifters.
Any imprecision can cause coherent spectral distortion of the realized unitary matrices and their subsequent samples.
The second source is processor noise, which induces small, random temporal fluctuations in the unitary matrices.
The measured intensities are, therefore, averages over this ensemble of slightly random unitarities.
This averaging mechanism inherently breaks unitarity and constitutes the primary source of decoherence in the device.
Next, we develop a phenomenological model to account for these effects.
This model will allow us to characterize the level of errors stemming from coherent spectral distortion and decoherence.

\subsection{Reconstruction in the presence of coherent spectral distortion and decoherence}\label{subsection:reconstruction_imperfect}

The idealized assumptions of the measurement process must be confronted with the physical realization of the unitary operations. In the experimental hardware, arbitrary unitary matrices are synthesized via a Clements decomposition~\cite{clements2016optimal}, which maps the target transformation $U$ onto an integrated mesh of Mach-Zehnder interferometers and phase shifters. The global transformation is the ordered product of these local operations, $U = \prod_{k=1}^N T_k(\phi_k)$, where $\phi_k$ are the programmable phases of the individual optical components.

During physical operation, the exact phase imparted by each component deviates from the ideal setting. We decompose this microscopic phase error into a time-independent systematic calibration offset and a zero-mean dynamic temporal fluctuation: $\delta\phi_k = \delta\phi_k^{\text{static}} + \delta\phi_k^{\text{noise}}$. We model the dynamic temporal noise as independent Gaussian random variables $\delta\phi_k^{\text{noise}} \sim \mathcal{N}(0, \sigma_k^2)$.

As detailed in Appendix~\ref{AppendixD}, integrating these microscopic fluctuations over the Haar measure rigorously separates the errors into two distinct macroscopic effects. The intended unitary transformation $U$ is physically realized as a perturbed product $UU_{c}$~\cite{strikis2023quantum}, where the static coherent spectral distortion $U_{c}$ represents the ensemble-averaged unitary deviation from the identity. This distortion modifies the measurement probabilities in the computational basis to
\begin{equation}
    P(b_i=1) = (UU_{c}\rho {U_{c}}^{\dagger}U^{\dagger})_{ii}.
\label{eq:prob_err}
\end{equation}

Simultaneously, the configuration-dependent temporal noise acting across the interferometric mesh is homogenized by the geometric symmetry of the Haar ensemble. This dynamic degradation converges strictly to a global depolarizing channel~\cite{Huang2020}:
\begin{equation}
    \rho'=(1-p)\rho +\frac{p}{d}\mathbb{I},
\end{equation}
where the parameter $p$ quantifies the severity of decoherence.
The combined impact of unitary imperfection and depolarization yields the effective measurement probabilities:
\begin{equation}
    P(b_i=1) = (1-p)(UU_{c}\rho {U_{c}}^{\dagger}U^{\dagger})_{ii}+\frac{p}{d}.\label{EMP}
\end{equation}
If we keep on applying the standard reconstruction procedure to these corrupted statistics, the resulting intermediate term becomes
\begin{equation}
    \tilde{\rho}_{ab}^{(1)} =(1-p)\sum_{i,j,k} U_{ai}^{*} U_{ij} (U_{c}\rho {U_{c}}^{\dagger})_{jk}U_{ik}^{*} U_{ib}+\frac{p}{d}\delta_{ab}.
\label{eq:rho_ab_err}
\end{equation}
Averaging over the Haar measure of the unitary group subsequently yields
\begin{equation}
    I_{ab}=(1-p)\frac{1}{d+1}\left[(U_{c}\rho {U_{c}}^{\dagger})_{ab}+\delta_{ab}\right]+\frac{p}{d}\delta_{ab}.
\end{equation}
Finally, the application of the reconstruction formula (\ref{eq:recform}) reveals the true nature of the recovered state:
\begin{equation}
    \tilde{\rho}_{ab}=(1-p)(U_{c}\rho {U_{c}}^{\dagger})_{ab}+\frac{p}{d}\delta_{ab}.\label{eq:imperfect}
\end{equation}
Equation (\ref{eq:imperfect}) reveals the mechanism of the crossover from a variance-dominated regime to a bias-dominated regime. The mathematical distribution governing the sampled unitary matrices remains the strictly invariant Haar measure $dU$. Instead, it is the effective measurement operator---and consequently the reconstructed state---that undergoes a physical transformation. The term $(1-p)(U_c \rho U_c^\dagger)$ represents a persistent rotation away from the target state, coupled with an isotropic contraction of the state vector due to depolarization. This combined effect locks the reconstruction into a permanent, shifted orbit, preventing convergence to the ideal target state even in the limit of infinite statistical accumulation.

\subsection{Spectral analysis of reconstructed density matrices}

\begin{figure}[htbp]
    \centering
    \includegraphics[width=0.58\linewidth]{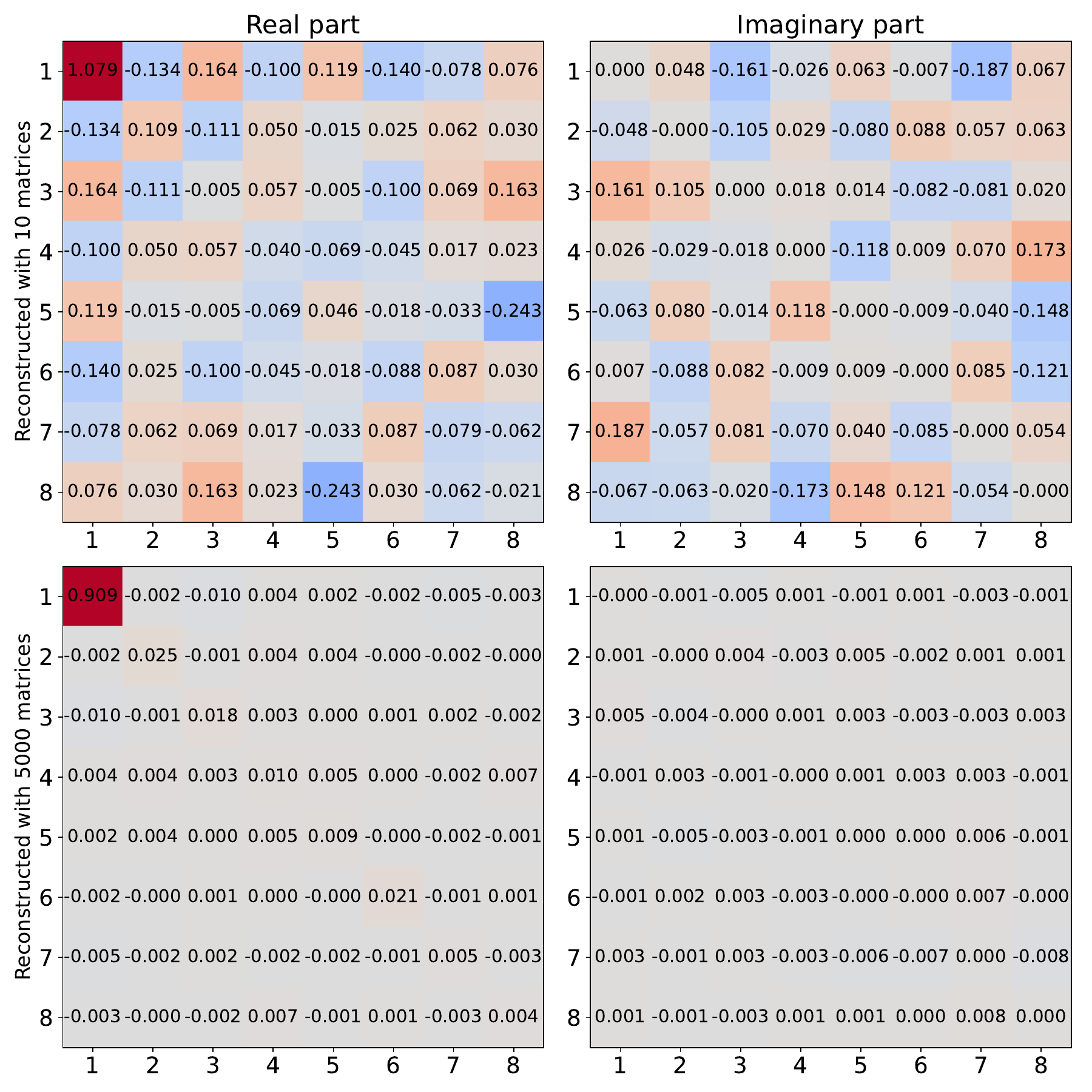}
    \includegraphics[width=0.39\linewidth]{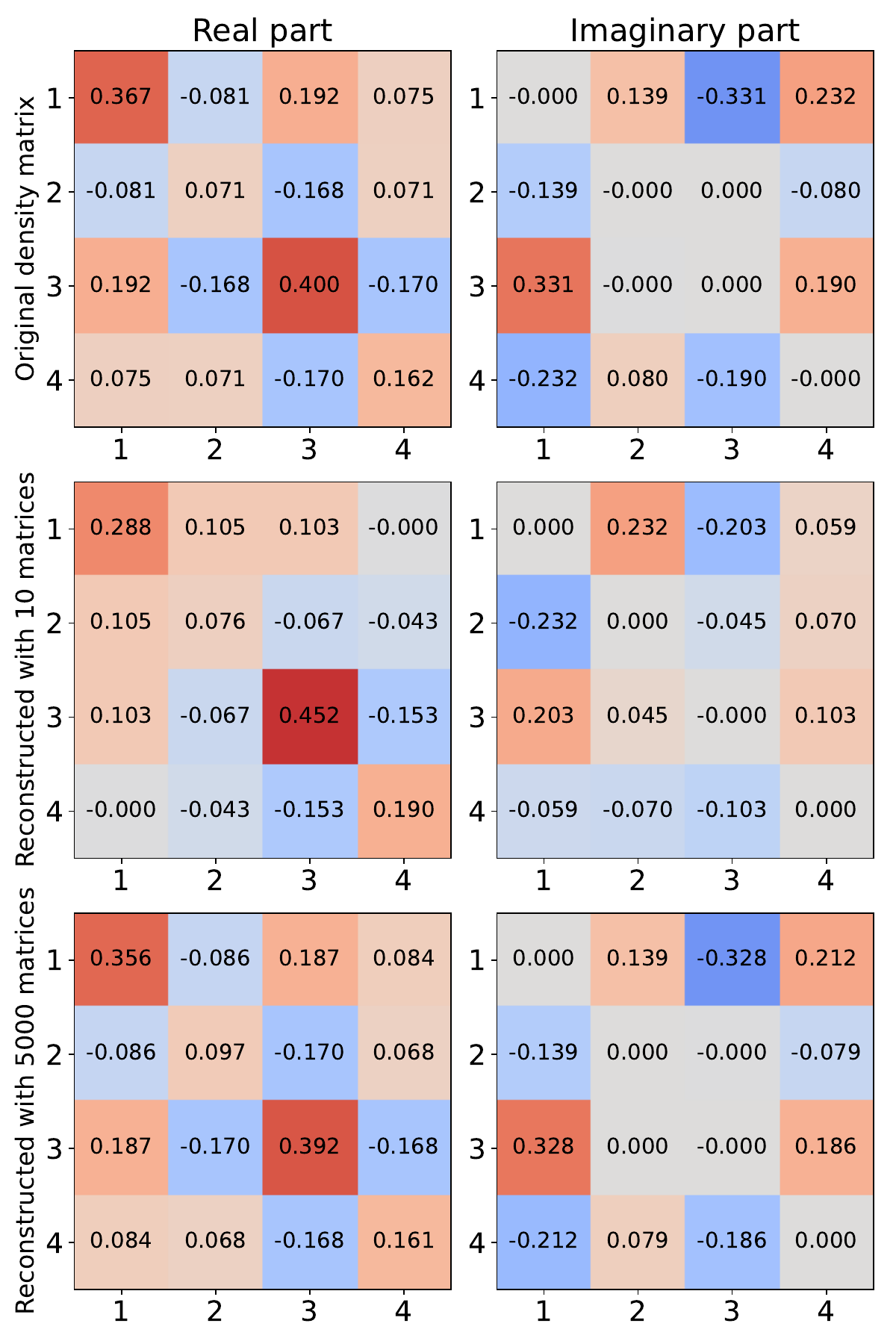}
    \caption{Experimental reconstruction of density matrices via the photonic processor.
\textbf{Left:} The density matrix of a trivial state in an 8-dimensional space.
\textbf{Right:} The density matrix of a non-trivial state in a 4-dimensional space. Colored squares in the heatmaps denote matrix elements;
real and imaginary parts are plotted separately with values indicated.
Reconstructions utilized up to 5000 unitary transformations.}
    \label{fig:expfig1ab}
\end{figure}

Our analysis begins with the relationship established in Eq.~(\ref{eq:imperfect}), which links the true quantum state to its reconstructed counterpart.
A significant theoretical challenge arises here: the reconstructed state $\tilde{\rho}$ is corrupted simultaneously by coherent spectral distortions ($U_{c}$) and the depolarizing channel ($p$).
Distinguishing between these two error sources is non-trivial, as one represents a reversible rotation of the basis, while the other represents an irreversible loss of information.
To resolve this ambiguity and isolate the error parameters, we exploit the spectral properties of the density matrix.
Since eigenvalues are invariant under unitary transformations, they are naturally immune to the coherent spectral distortion inherent in $U_{c}$.
By performing an eigendecomposition of the original density matrix $\rho=W\Lambda W^\dagger$ where $W$ is the unitary matrix of eigenvectors and $\Lambda$ is the diagonal matrix of eigenvalues $\lambda_i$ we can derive a clear structural representation of the noise.
Substituting this decomposition into Eq.~(\ref{eq:imperfect}) and applying the identity $U_{c}WW^\dagger {U_{c}}^{\dagger}=\mathbb{I}$ reveals the simplified form:
\begin{equation}
    \tilde{\rho}=U_{c}W D W^\dagger {U_{c}}^{\dagger},
\end{equation}
where the diagonal matrix $D$ is given by
\begin{equation}
    D=(1-p)\Lambda +\frac{p}{d}\mathbb{I}.
\end{equation}
The eigenvalues of the reconstructed state, $D_i=(1-p)\lambda_i+p/d$, effectively capture the interplay between the ideal state's spectrum and the depolarizing channel.
This theoretical framework provides a direct method for extracting $p$, independent of coherent spectral distortion.
It is important to define the scope and limitations of this phenomenological model.
While we use the term ``Hardware Horizon'' to describe the observed saturation of the reconstruction error, this threshold represents a device-dependent error floor rather than an immutable law of physics.
Modeling hardware imperfections via a single, fixed unitary distortion $U_c$ is an effective, ensemble-averaged approximation over the Haar measure.
In physical reality, distortions within the photonic chip inherently depend on the specific dynamic phase settings of the unit cells for each applied unitary operation.
Consequently, $U_c$ functions as a statistical abstraction that captures the mean coherent deviation of the processor, rather than a strict, universally static hardware constant.
However, a practical complication remains.
Experimental reconstructions are derived from finite datasets, introducing statistical fluctuations.
This noise disproportionately affects the smaller eigenvalues, rendering the lower end of the spectrum unreliable for precise estimation.
To mitigate this, we restrict our analysis to the most robust component of the spectrum: the leading eigenvalue.
As this value is least susceptible to statistical noise, it offers the most dependable basis for calculating the error parameter.
Using the leading eigenvalue, we approximate the depolarizing channel parameter as:
\begin{equation}
    p\approx \frac{\lambda_1-D_1}{\lambda_1-1/d}.
\label{eq:depolar_general}
\end{equation}
In the context of our measurement protocols, the original density matrix represented a pure state $\rho=\ket{\psi}\bra{\psi}$.
Consequently, the theoretical spectrum consists of a single dominant eigenvalue $\lambda_1=1$, with the remaining $d-1$ eigenvalues vanishing.
Under these conditions, the estimator simplifies to:
\begin{equation}
    p\approx (1-D_1)\frac{d-1}{d}.
\label{eq:depolar_pure}
\end{equation}
The eigenvalues obtained from the experimentally reconstructed density matrices are presented in Table~\ref{tab:eigenvalues}.

\begin{table*}[htbp]
    \centering
    \caption{Eigenvalues of the experimentally reconstructed density matrices for the measurement protocols detailed in Section~\ref{sec:protocols}, utilizing $M=5000$ unitary matrices for reconstruction.}
    \label{tab:eigenvalues}
    \begin{ruledtabular}
    \begin{tabular}{ccccccccc}
        \textbf{Protocol} & \textbf{1} & \textbf{2} & \textbf{3} & \textbf{4} & \textbf{5} & \textbf{6} & \textbf{7} & \textbf{8} \\
        \hline
        \textbf{I.}   &  0.90889 & -0.00820 & -0.00075 & 0.03102 & 0.02638 & 0.02146 & 0.01411 & 0.00707 \\
        \textbf{II.}  &  0.90105 & 0.04071 & 0.02831 & 0.01643 & 0.01461 & 0.00754 & 0.00171 & -0.01038\\
        \textbf{III.} &  0.95372 & 0.00669 & 0.02906 & 0.01958 & -- & -- & -- & -- \\
        \textbf{IV.}  &  0.98087 & 0.02503 & 0.01116 & -0.01150 & -- & -- & -- & -- \\
    \end{tabular}
    \end{ruledtabular}
\end{table*}

Based on the dominant eigenvalues, we estimate the depolarizing channel parameters summarized in Table~\ref{tab:errors}. It is notable that the depolarization parameter is significantly higher for the $8\times 8$ matrices compared to the $4 \times 4$ cases.
This discrepancy arises because the larger matrices necessitate the involvement of more unit cells, thereby compounding the depolarization effect of individual cells within the measurement results.

\subsection{Frobenius norm and the isolation of coherent spectral distortion}

Figure~\ref{fig:expfig2} unveils a critical limitation in the reconstruction process: simply increasing the ensemble size $M$ of Haar random unitaries eventually ceases to yield accuracy gains.
Instead of converging to zero, the Frobenius norm arrests at a saturation plateau.
This persistent residual error compels us to ask: what factors prevent perfect reconstruction, even in the limit of infinite sampling?
We resolve this by dissecting the Mean Squared Error (MSE).
This phenomenological framework allows us to isolate transient statistical fluctuations from the systematic errors induced by decoherence and coherent spectral distortion:
\begin{equation}
    \mathbb{E}\left[ \| \tilde{\rho}^{(M)}-\rho\|_F^2 \right] =\mathbb{E}\left[ \| (\tilde{\rho}^{(M)}-\tilde{\rho})+(\tilde{\rho}-\rho)\|_F^2 \right].
\end{equation}
Here, $\tilde{\rho}^{(M)}$ represents the reconstruction from $M$ unitaries, while $\tilde{\rho}$ denotes the physically measured, distorted state (as defined in Eq.~\ref{eq:imperfect}), and $\rho$ is the ideal input state.
Since $\tilde{\rho}^{(M)}$ acts as an unbiased estimator of $\tilde{\rho}$, the cross-terms vanish ($\mathbb{E}[ \tilde{\rho}^{(M)}-\tilde{\rho}]=0$), allowing us to separate the MSE into two distinct terms:
\begin{equation}
    \mathbb{E}\left[ \| \tilde{\rho}^{(M)}-\rho\|_F^2 \right] =\mathbb{E}\left[ \| \tilde{\rho}^{(M)}-\tilde{\rho}\|^2_F\right]+\mathbb{E}\left[ \|\tilde{\rho}-\rho\|^2_F \right].\label{eq:MSE}
\end{equation}
The first term represents the statistical error of estimation.
As derived in Eq.~(\ref{eq:Frobeniuserrgen}), this error scales inversely with the sample size:
\begin{equation}
    \mathbb{E}\left[ \| \tilde{\rho}^{(M)}-\tilde{\rho}\|^2_F\right]=\frac{d\, \text{Tr}(\tilde{\rho}^2)-1}{M}.
\end{equation}
Substituting our distortion model (\ref{eq:imperfect}), and considering our experimental protocol where the input $\rho=\ket{\psi}\bra{\psi}$ is a pure state, this simplifies to a vanishing contribution:
\begin{equation}
     \mathbb{E}\left[ \| \tilde{\rho}^{(M)}-\tilde{\rho}\|^2_F\right]=(1-p)^2\frac{d-1}{M}.
\end{equation}
As $M$ increases, this term disappears. Yet, the saturation observed in Figure~\ref{fig:expfig2} remains.
This brings us to the second term in Eq.~(\ref{eq:MSE}) the dominant factor responsible for the leveling off of the Frobenius norm.
Independent of measurement quantity, this term contains the structural imperfections of the system:
\begin{widetext}
\begin{equation}
 \mathbb{E}\left[ \|\tilde{\rho}-\rho\|^2_F \right]=   p^2\left[\text{Tr}(\rho^2)-\frac{1}{d}\right]+2(1-p)\left[\text{Tr}(\rho^2)-\text{Tr}(U_c\rho U_c^\dagger \rho)\right].
\end{equation}
\end{widetext}
Here we can isolate the decoherence contribution (proportional to $p^2$) and the coherent spectral distortion
\begin{equation}
    \mathcal{E}_c=\text{Tr}(\rho^2)-\text{Tr}(U_c\rho U_c^\dagger \rho).
\end{equation}
For our pure state protocol, the total systematic error crystallizes into:
\begin{equation}
     \mathbb{E}\left[ \|\tilde{\rho}-\rho\|^2_F \right]=   p^2\left(1-\frac{1}{d}\right)+2(1-p)\left[1-|\bra{\psi}U_c\ket{\psi}|^2\right].\label{slope}
\end{equation}
This result offers a direct probe into the unitary error matrix $U_c$.
By working in a basis where $\ket{\psi}$ aligns with the first eigenbasis vector $\ket{100\dots}$, the term $\bra{\psi}U_c\ket{\psi}$ corresponds to the diagonal element $\{U_c\}_{11}$.
Utilizing the unitarity condition $\sum_j^d|\{U_c\}_{1j}|^2=1$, we arrive at a precise definition of the coherent spectral distortion in terms of the matrix's off-diagonal elements:
\begin{equation}
    \mathcal{E}_c=1-|\bra{\psi}U_c\ket{\psi}|^2=\sum_{j\neq 1}^d|\{U_c\}_{1j}|^2.\label{saturation}
\end{equation}

Consequently, the Hardware Horizon is not merely an accuracy limit; it is a robust spectral fingerprint of the specific processor. Equation (\ref{saturation}) demonstrates that the saturation level is invariant under unitary randomization, depending solely on the mean off-diagonal elements of the coherent distortion matrix $U_c$.
Consequently, the saturation of the Frobenius norm is not merely a limit to accuracy; it is a diagnostic tool.
It reveals that the coherent spectral distortion is exactly the sum of squares of the off-diagonal ($j\neq 1$) elements of $U_c$.
The mean square of the off-diagonal elements is thus given by:
\begin{equation}
\epsilon^2=\mathbb{E}\left[ |\{U_c\}_{1j}|^2\right]=\frac{1}{d-1}\sum_{j\neq 1}^d|\{U_c\}_{1j}|^2=\frac{\mathcal{E}_c}{d-1},
\end{equation}
where $\epsilon$ is the average magnitude of the coherent distortion matrix elements.

\begin{figure}[htbp]
    \centering
    \includegraphics[width=0.49\linewidth]{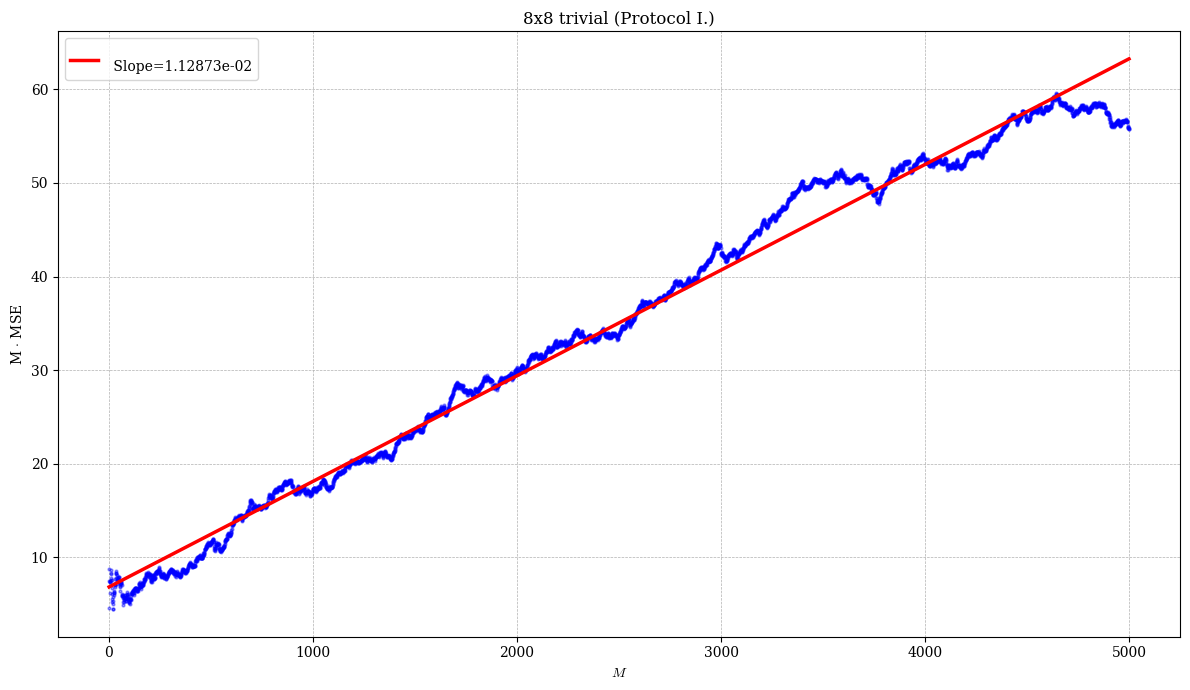}
    \hfill
    \includegraphics[width=0.49\linewidth]{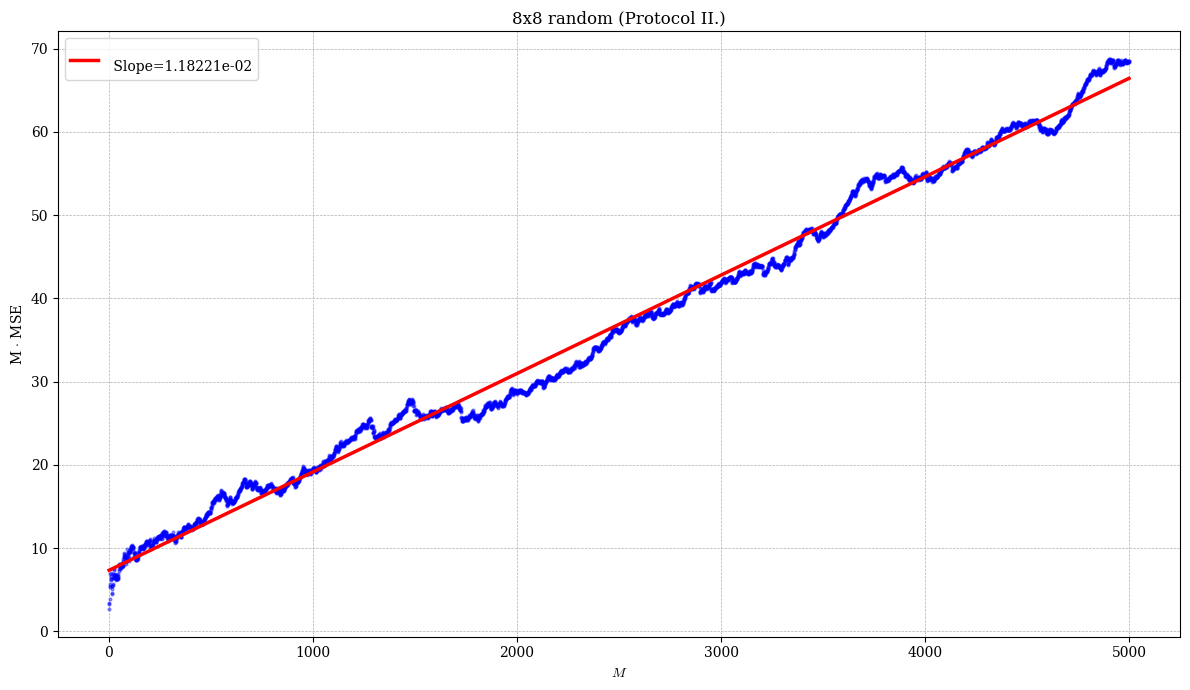}
    
    \vspace{0.5em} 
    
    \includegraphics[width=0.49\linewidth]{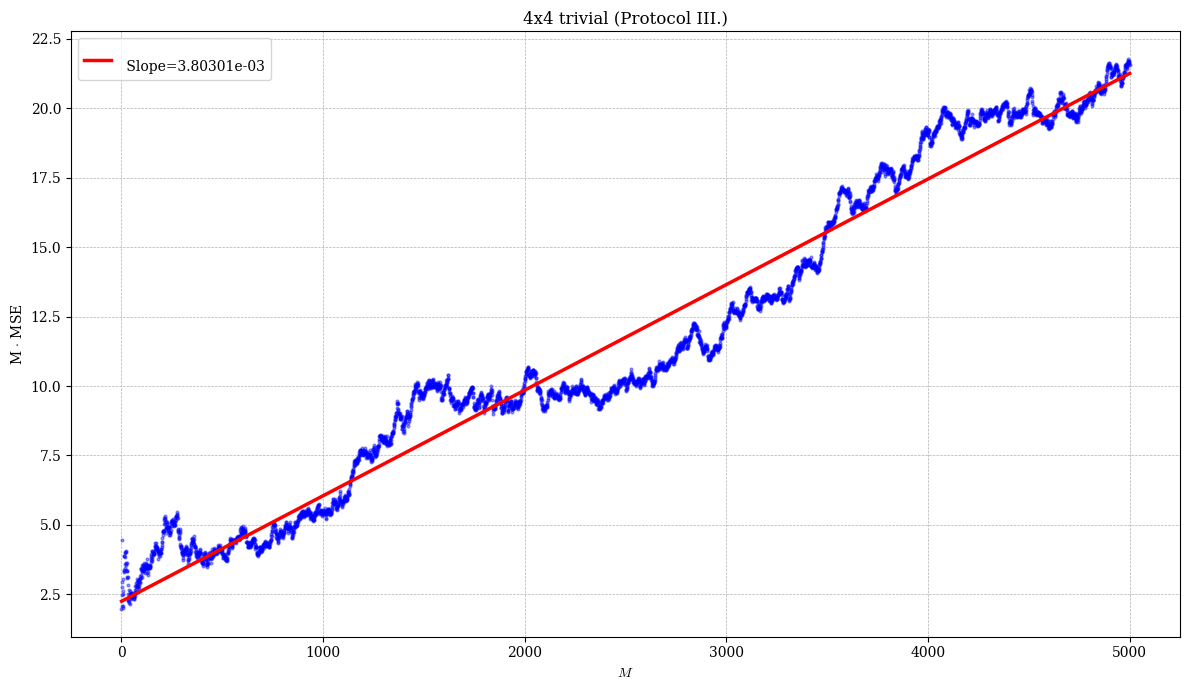}
    \hfill
    \includegraphics[width=0.49\linewidth]{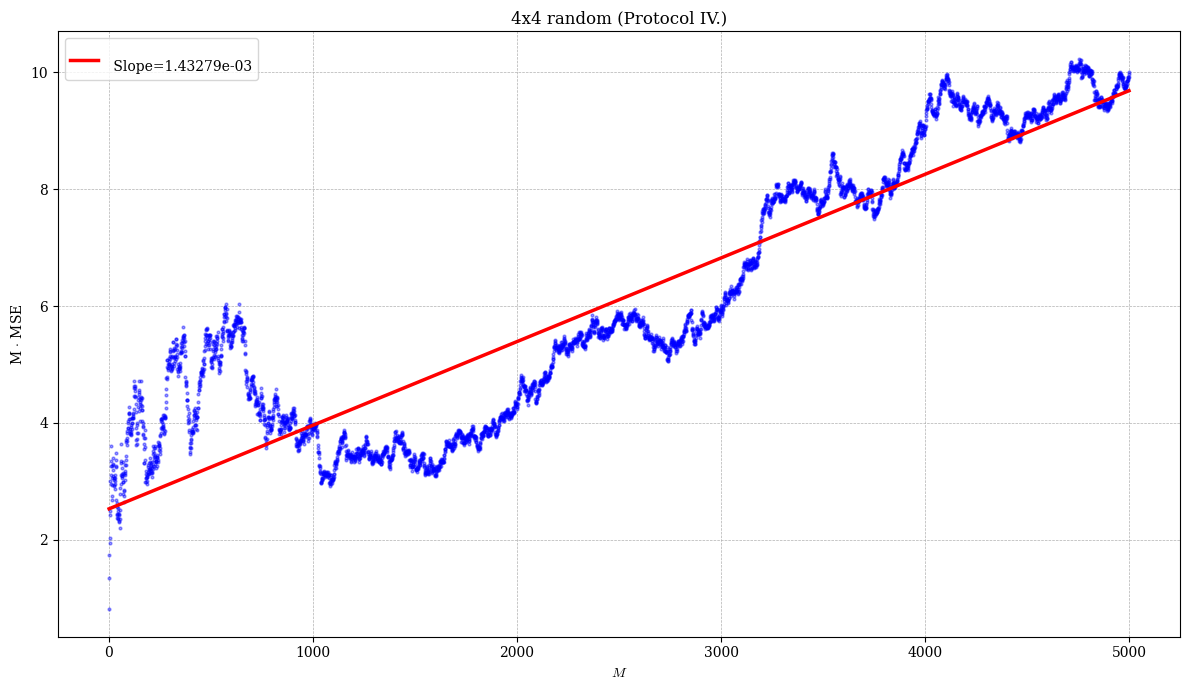}

\caption{\textbf{Robust spectral fingerprint of the photonic processor.} 
The scaled reconstruction error $M||\tilde{\rho}^{(M)} - \rho||_F^2$ as a function of sample size $M$ for the four experimental protocols.
The strict linearity confirms the validity of the phenomenological error model (Eq.~\ref{slope}), separating statistical fluctuations from systematic hardware distortion.
While the slopes vary significantly due to state-dependent decoherence ($p$), the underlying coherent spectral distortion extracted from these slopes remains highly consistent.
In all regimes, the static unitary distortion converges to a mean off-diagonal element magnitude of $\epsilon \approx 0.012$ (see Table~\ref{tab:errors}), establishing this value as the operational \textit{Hardware Horizon} of the device.
The panels correspond to Protocol I (top left) $Slope= 1.12873 \cdot 10^{-2} \pm 1.82596 \cdot 10^{-5}$, Protocol II (top right) $Slope=1.18221 \cdot 10^{-2} \pm 1.53545 \cdot 10^{-5}$, Protocol III (bottom left) $Slope=3.80301  \cdot 10^{-3} \pm  1.06298 \cdot 10^{-5}$, and Protocol IV (bottom right) $Slope=1.43278 \cdot 10^{-3} \pm 9.30978 \cdot 10^{-6}$.\label{fig:U}}
\end{figure}

\begin{table}[htbp]
\centering
\caption{Depolarizing channel parameters calculated from the largest eigenvalues in Table~\ref{tab:eigenvalues} via $p\approx (1-D_1)\frac{d-1}{d}$ (first column) and the average magnitude of the coherent distortion matrix elements $\epsilon=\left[\frac{Slope-p^2\left(1-\frac{1}{d}\right)}{2(1-p)(d-1)}\right]^{1/2}$ calculated from the slopes in Fig.~\ref{fig:U} (second column).}
\label{tab:errors}
\begin{ruledtabular}
\begin{tabular}{ccc}
\textbf{Protocol}&\textbf{$p$} & \textbf{$\epsilon$}  \\
\hline
\textbf{I.}& 0.10412 & 0.01198 \\
\textbf{II.}& 0.11308&0.01164 \\
\textbf{III.}& 0.06170 &0.01297 \\
\textbf{IV.}& 0.02550& 0.01271 \\
\end{tabular}
\end{ruledtabular}
\end{table}

\subsection{Consistency of the scaling law}

To isolate the true coherent spectral distortion from the noise, we examine the structure of the MSE curves.
By plotting the scaled error $M||\tilde{\rho}^{(M)} - \rho||_F^2$ as a function of $M$, a linear relationship emerges (Fig.~\ref{fig:U}), characterized by the slope defined in Eq.~\ref{slope}.
This linearity allows us to pierce through the measurement uncertainties.
Utilizing our established estimates for $p$ (Table~\ref{tab:errors}), we determine the magnitude of the coherent distortion, $\epsilon$, via the relation:
\begin{equation}
    \epsilon = \left[ \frac{\text{Slope} - p^2(1 - 1/d)}{2(1-p)(d-1)} \right]^{1/2}.
\end{equation}

The visualization of these slopes in Fig.~\ref{fig:U} and the calculated average magnitudes in Table~\ref{tab:errors} reveal consistency.
While the observed slopes and depolarization parameters $p$ fluctuate by orders of magnitude---driven by the differing complexity of the target states---the derived coherent spectral distortion exhibits a rigid invariance.
In all scenarios, the static unitary distortion converges to a mean off-diagonal element magnitude of $\epsilon \approx 0.012$.
This consistency confirms that the Hardware Horizon is not an artifact of specific measurement protocols but a robust, state-independent spectral fingerprint of the specific photonic architecture.
Consequently, $\epsilon$ represents the effective ``resolution limit'' of the hardware, defining the boundary where device physics inevitably supersedes statistical scaling.

\section{Conclusion}
\label{sec:conclusion}

In this work, we have experimentally established the existence of an operational ``Hardware Horizon''---a hardware-induced systematic error floor that strictly limits the efficacy of classical shadow tomography on integrated photonic processors.
By pushing the measurement count $M$ beyond the standard statistical limits, we observed a distinct crossover where the theoretical $\mathcal{O}(M^{-1/2})$ scaling collapses into a saturation regime governed entirely by device physics, marking a transition from a variance-dominated regime to a bias-dominated regime.
This discovery reframes the challenge of NISQ-era tomography. We have demonstrated that the primary obstacle to high-fidelity reconstruction is not the acquisition of statistics, but the spectral distortion of the unitary implementation itself.
Our phenomenological model reveals that this horizon is defined by a specific interplay between static coherent spectral distortion ($\epsilon$) and dynamic decoherence ($p$), creating a device-specific error floor that no amount of additional sampling can overcome.
Consequently, the standard promise of shadow tomography polynomial resource scaling is effectively voided once the system reaches this operational hardware limit.
This realization creates an urgent imperative for the field: to move beyond passive error characterization and toward active spectral compensation.
The open challenge is no longer just about measuring faster, but about whether we can develop algorithmic inversions that ``learn'' the distorted unitary measure in real time.
To breach this horizon, we propose the development of structure-aware reconstruction kernels that abandon the assumption of an ideal Haar ensemble.
Instead, by incorporating the experimentally learned distortion matrix $U_c$ characterized via independent transfer-matrix inversion or efficient process tomography directly into the Weingarten integration, it becomes possible to convert static coherent spectral distortion from a device-dependent saturation limit into a corrigible systematic bias, effectively extending the reach of shadow tomography into the sub-horizon regime.

\begin{acknowledgments}
This research was carried out with the support of the Ministry of Culture and Innovation, funded by the National Research Development and Innovation Fund, under project number 2022-2.1.1-NL-2022-00004.
\end{acknowledgments}

\appendix

\section{Density matrix reconstruction}\label{subsec:reconstruction_theory}
For a $d$-dimensional quantum system, let the input state be described by the density matrix $\rho$.
Suppose a unitary transformation $U$ is applied to this state.
In that case, the probability of observing an outcome $b_i$ (e.g., detecting a particle in the $i$-th output channel of a photonic circuit) is given by the Born rule.
For instance, if we are interested in the diagonal elements of the transformed state $U\rho U^{\dagger}$, which correspond to measurement probabilities in the computational basis, we have:
\begin{equation}
      P(b_i=1) = (U\rho U^{\dagger})_{ii} = \sum_{j,k} U_{ij}\rho_{jk}U_{ik}^{*} = \rho_{ii}^{exp}
    \label{eq:prob_bi_thesis_2}
\end{equation}
where $U_{ik}^{*}$ is the complex conjugate of $U_{ik}$.
From a single measurement (or a set of measurements yielding probabilities $P(b_i=1)$ for a fixed $U$), we can form a diagonal matrix, $\rho^{exp}$, where $\rho_{ii}^{exp} = P(b_i=1)$ as stated in \ref{eq:prob_bi_thesis_2}.
An estimate of the original density matrix element $\rho_{ab}$ after one such measurement procedure (applying $U$ and measuring probabilities) can be constructed as:
\begin{equation}
    \hat{\rho}_{ab}^{(1)} = \sum_{i} U_{ai}^{*} \rho_{ii}^{exp} U_{ib} = \sum_{i} U_{ai}^{*} \left( \sum_{j,k} U_{ij}\rho_{jk}U_{ik}^{*} \right) U_{ib}.
\label{eq:rho_ab_single_U_thesis_5}
\end{equation}
Rearranging this expression, we get:
\begin{equation}
    \hat{\rho}_{ab}^{(1)} = \sum_{i,j,k} U_{ai}^{*} U_{ij} \rho_{jk} U_{ik}^{*} U_{ib}.
\label{eq:rho_ab_rearranged_thesis_6}
\end{equation}
To obtain a more accurate reconstruction of the original density matrix $\rho$, we must average the results obtained from many different unitary transformations $U$, drawn from a suitable distribution, in our case the Haar distribution.
Let $I_{ab}$ be the quantity obtained by averaging $\hat{\rho}_{ab}^{(1)}$ over the unitary group with respect to the Haar measure $dU$:
\begin{equation}
    I_{ab} = \mathbb{E}_{U}[\hat{\rho}_{ab}^{(1)}] = \int dU \sum_{i,j,k} U_{ai}^{*} U_{ij} \rho_{jk} U_{ik}^{*} U_{ib}.
\label{eq:I_ab_integral_thesis_7}
\end{equation}
The evaluation of this integral involves terms of the form $\int dU U_{a'b'} U_{c'd'} U_{e'f'}^{*} U_{g'h'}^{*}$, which are known as Weingarten functions~\cite{Collins2022weingarten}.
The general formula for calculating such integrals is given by:
\begin{widetext}
\begin{equation}
\int_{U_d} U_{i_1 j_1} \cdots U_{i_q j_q} U_{i_1' j_1'}^* \cdots U_{i_q' j_q'}^* dU  =\sum_{\sigma, \tau \in S_q} \delta_{i_1' i_{\sigma(1)}} \cdots \delta_{i_q' i_{\sigma(q)}} \delta_{j_1 j_{\tau(1)}} \cdots \delta_{j_q j_{\tau(q)}} Wg(\sigma \tau^{-1}, d)
    \label{eq:Wg_1}
\end{equation}
\end{widetext}
where $Wg(\sigma, d)$ denotes the Weingarten functions:
\begin{equation}
    Wg(\sigma, d) = \frac{1}{q!^2} \sum_\lambda \frac{\chi^\lambda(e)^2 \chi^\lambda(\sigma)}{s_{\lambda, d}(1)},
    \label{eq:Wg_2}
\end{equation}
where $d$ is the dimension of our unitary matrices.
In our case for equation \ref{eq:rho_ab_rearranged_thesis_6} the general formula is given by:
\begin{widetext}
\begin{equation}
    \begin{aligned}
        \int dU \, U_{i_1 j_1} U_{i_2 j_2} U_{i_1' j_1'}^* U_{i_2' j_2'}^* &= (\delta_{i_1 i_1'} \delta_{i_2 i_2'} \delta_{j_1 j_1'} \delta_{j_2 j_2'} + \delta_{i_1 i_2'} \delta_{i_2 i_1'} \delta_{j_1 j_2'} \delta_{j_2 j_1'}) Wg(1^2, d) \\
        &\quad - (\delta_{i_1 i_1'} \delta_{i_2 i_2'} \delta_{j_1 j_2'} \delta_{j_2 j_1'} + \delta_{i_1 i_2'} \delta_{i_2 i_1'} \delta_{j_1 j_1'} \delta_{j_2 j_2'}) Wg(2, d),
    \end{aligned}
    \label{eq:Wg_3}
\end{equation}
\end{widetext}
where:
\begin{equation}
    \begin{aligned}
      Wg(1^2, d) = \frac{1}{d^2 - 1}, \quad \quad Wg(2, d) = \frac{-1}{d(d^2 - 1)},
    \end{aligned}
    \label{eq:Wg_4}
\end{equation}
and in result:
\begin{widetext}
\begin{equation}
        \int dU \, U_{i_1 j_1} U_{i_2 j_2} U_{i_1' j_1'}^* U_{i_2' j_2'}^* = \frac{\delta_{i_1 i_1'} \delta_{i_2 i_2'} \delta_{j_1 j_1'} \delta_{j_2 j_2'} + \delta_{i_1 i_2'} \delta_{i_2 i_1'} \delta_{j_1 j_2'} \delta_{j_2 j_1'}}{d^2 - 1} 
        \quad - \frac{\delta_{i_1 i_1'} \delta_{i_2 i_2'} \delta_{j_1 j_2'} \delta_{j_2 j_1'} + \delta_{i_1 i_2'} \delta_{i_2 i_1'} \delta_{j_1 j_1'} \delta_{j_2 j_2'}}{d(d^2 - 1)}.
\label{eq:Wg_5}
\end{equation}
\end{widetext}
The symmetries of the problem allow us to rearrange the indices in equation \ref{eq:Wg_5} and arrive at the arrangement of \ref{eq:I_ab_integral_thesis_7} eventually getting the exact solution to the integral in question:
\begin{equation}
    \begin{aligned}
        \int dU U_{ia}^* U_{ij} U_{ik}^* U_{ib} &=  \frac{1}{d(d+1)} (\delta_{ja} \delta_{bk} + \delta_{jk} \delta_{ba}).
\end{aligned}
    \label{eq:Wg_7}
\end{equation}

The result for the specific fourth-order moment in Eq.~\eqref{eq:I_ab_integral_thesis_7}, after summing over index $i$ (which introduces a factor of $d$, the dimension of the Hilbert space), the integral simplifies significantly.
The relevant integral is:
\begin{equation}
    \sum_{i} \int dU U_{ai}^{*} U_{ij} U_{ik}^{*} U_{ib} = \frac{1}{d+1}(\delta_{aj}\delta_{kb} + \delta_{ak}\delta_{jb}).
\end{equation}
Summing over $i$ in Eq.~\eqref{eq:I_ab_integral_thesis_7} using this result (and re-indexing summation variables for clarity) yields:
\begin{equation}
    I_{ab} = \sum_{j,k} \frac{1}{d+1}(\delta_{aj}\delta_{kb} + \delta_{ak}\delta_{jb}) \rho_{jk}.
\end{equation}
This simplifies to:
\begin{equation}
    I_{ab} = \frac{1}{d+1} (\rho_{ab} + \delta_{ab} \text{Tr}(\rho)).
\label{eq:I_ab_final_thesis_16}
\end{equation}
For a physical density matrix, the trace $\text{Tr}(\rho) = 1$.
Therefore, we can express the elements of the original density matrix $\rho_{ab}$ in terms of the averaged quantity $I_{ab}$:
\begin{equation}
    \rho_{ab} = (d+1)I_{ab} - \delta_{ab}.
\label{eq:rho_reconstructed_final_thesis_17}
\end{equation}
In practice, $I_{ab}$ is estimated by averaging $\rho_{ab}^{(k)} = \sum_{i} U_{ai}^{*} \rho_{ii}^{exp} U_{ib}$ (from Eq.~\eqref{eq:rho_ab_rearranged_thesis_6} for each unitary $U^{(k)}$) over a finite number $M$ of unitary transformations randomly chosen from a Haar distribution:

\begin{equation}
    \overline{I_{ab}}^{(M)} = \frac{1}{M}\sum_{k=1}^{M}\rho_{ab}^{(k)} = \mathbb{E}_{U}^{(M)}\left[\sum_i U_{ai}^{\dagger} \rho_{ii}^{\text{exp}} U_{ib}\right].
\label{eq:rho_reconstructed_final_thesis_18}
\end{equation}
The reconstructed matrix is then:
\begin{equation}
    \rho_{ab}^{(rec)}=(d+1)\overline{I_{ab}}^{(M)}-\delta_{ab}.
    \label{eq:rho_reconstructed_final_thesis_19}
\end{equation}
The accuracy of this reconstruction depends on $M$ and the dimension $d$, later shown in Appendix~\ref{app:frobenius}.

\section{Generation of Haar measure unitary matrices}
\label{app:haar_measure}
The reconstruction formalism detailed in Appendix~\ref{sec:theory}, particularly the use of Weingarten functions for averaging, mandates that the unitary transformations $U$ are drawn from a uniform distribution over the unitary group, according to the Haar measure.
To generate such $d \times d$ unitary matrices, we employ the following standard algorithm~\cite{haarinvar}:
\begin{enumerate}
    \item Generate an $d \times d$ complex matrix $Z$, where each element $Z_{ij} = x_{ij} + iy_{ij}$.
The real part $x_{ij}$ and imaginary part $y_{ij}$ are independent random variables drawn from a standard normal distribution $\mathcal{N}(0,1)$.
    \item Perform a QR decomposition of $Z$ such that $Z = QR$, where $Q$ is a unitary matrix and $R$ is an upper triangular matrix.
    \item Construct a diagonal matrix $D$ where each diagonal element $D_{kk}$ is the phase of the corresponding diagonal element of $R$: $D_{kk} = R_{kk} / |R_{kk}|$ for $R_{kk} \neq 0$, and $D_{kk}=1$ if $R_{kk}=0$.
    \item The resulting unitary matrix $U = QD$ is then distributed according to the Haar measure.
\end{enumerate}

\section{Variance of reconstruction}\label{app:frobenius}

In the photonic implementation of the shadow tomography, we do not obtain a single discrete outcome.
Instead, we measure the intensities at all $d$ output ports simultaneously.
For a given Haar-random unitary $U$ acting on the state $\rho$, the probability of detecting a photon in port $k$ is given by the diagonal element of the rotated density matrix:
\begin{equation}
    p_k = \bra{k} U \rho U^\dagger \ket{k} = (U \rho U^\dagger)_{kk}.
\end{equation}
We construct the single-snapshot estimator $\hat{\rho}$ using the full distribution $\{p_k\}_{k=1}^d$:
\begin{equation}
    \hat{\rho} = (d+1) \sum_{k=1}^d p_k U^\dagger \ket{k}\bra{k} U - \mathbb{I} = (d+1) A - \mathbb{I},
\end{equation}
where we have defined the matrix $A = U^\dagger \text{diag}(p) U$.

\subsection{Expectation value (first moment)}
We first verify that the estimator is unbiased.
The matrix $A$ can be written as:
\begin{equation}
    A = \sum_{k=1}^d \bra{k} U \rho U^\dagger \ket{k} U^\dagger \ket{k}\bra{k} U.
\end{equation}
Averaging over the Haar measure $dU$:
\begin{equation}
    \mathbb{E}[A] = \int dU \sum_k \bra{k} U \rho U^\dagger \ket{k} U^\dagger \ket{k}\bra{k} U = \mathcal{D}(\rho),
\end{equation}
where $\mathcal{D}(\rho) = \frac{\rho + \mathbb{I}}{d+1}$.
Thus:
\begin{equation}
    \mathbb{E}[\hat{\rho}] = (d+1)\left( \frac{\rho + \mathbb{I}}{d+1} \right) - \mathbb{I} = \rho.
\end{equation}

\subsection{Total Reconstruction error (sum of matrix elements)}
To find the global error, we calculate the Mean Squared Error (MSE) in the Frobenius norm for a single snapshot.
Since the estimator is unbiased, the MSE is the trace of the covariance:
\begin{equation}
    \text{MSE} = \mathbb{E}\left[ \| \hat{\rho} - \rho \|_F^2 \right] = \mathbb{E}\left[ \text{Tr}(\hat{\rho}^2) \right] - \text{Tr}(\rho^2).
\end{equation}

First, we compute $\text{Tr}(\hat{\rho}^2)$:
\begin{align}
    \hat{\rho}^2 &= \left[ (d+1) A - \mathbb{I} \right]^2 = (d+1)^2 A^2 - 2(d+1)A + \mathbb{I}. \\
    \text{Tr}(\hat{\rho}^2) &= (d+1)^2 \text{Tr}(A^2) - 2(d+1)\text{Tr}(A) + d.
\end{align}
Note that $\text{Tr}(A) = \sum_k p_k = 1$.
Now we analyze $A^2$:
\begin{equation}
    A^2 = \left( \sum_k p_k U^\dagger \ket{k}\bra{k} U \right) \left( \sum_j p_j U^\dagger \ket{j}\bra{j} U \right).
\end{equation}
Using the orthogonality $U^\dagger \ket{k}\bra{k} U U^\dagger \ket{j}\bra{j} U = \delta_{kj} U^\dagger \ket{k}\bra{k} U$, we simplify:
\begin{equation}
    A^2 = \sum_k p_k^2 U^\dagger \ket{k}\bra{k} U \implies \text{Tr}(A^2) = \sum_{k=1}^d p_k^2.
\end{equation}
Thus, the problem reduces to finding the average purity of the output distribution:
\begin{equation}
    \mathbb{E}[\text{Tr}(A^2)] = \mathbb{E}\left[ \sum_k (\bra{k} U \rho U^\dagger \ket{k})^2 \right].
\end{equation}
Using standard Weingarten calculus for the second moment of Haar random unitary entries, for any state $\rho$:
\begin{equation}
    \mathbb{E}\left[ \sum_k p_k^2 \right] = \frac{1 + \text{Tr}(\rho^2)}{d+1}.
\end{equation}
Substituting this back into the trace equation:
\begin{align}
    \mathbb{E}[\text{Tr}(\hat{\rho}^2)] = (d+1)\text{Tr}(\rho^2) - 1.
\end{align}
Finally, the MSE for a single snapshot is:
\begin{equation}
    \text{MSE} = [(d+1)\text{Tr}(\rho^2) - 1] - \text{Tr}(\rho^2) = d\, \text{Tr}(\rho^2) - 1.
\end{equation}
For $M$ snapshots, the variance scales by $1/M$.
\begin{equation}
    \mathbb{E}\left[ \| \hat{\rho}^{(M)} - \rho \|_F^2 \right] = \frac{d\, \text{Tr}(\rho^2) - 1}{M}.
\end{equation}
\textbf{Special Case (Pure State):} If $\rho$ is pure ($\text{Tr}(\rho^2)=1$), this simplifies to:
\begin{equation}
    \text{MSE} = \frac{d - 1}{M}.
\end{equation}

This $1/M$ variance characterizes the Statistical Regime. The failure of this term to vanish in the experimental data (Fig.~\ref{fig:expfig2}) constitutes the mathematical definition of the Hardware Horizon.

\section{Derivation of the effective measurement probabilities from heterogeneous statistical noise} \label{AppendixD}

To establish the physical origin of the depolarizing channel and the effective measurement probability expression (Eq.~\ref{EMP}), we analyze the effect of dynamic statistical noise traversing the integrated photonic processor. Let the ideal target unitary matrix be synthesized via the Clements decomposition\cite{clements2016optimal} as the ordered product of local transformations:
\begin{equation}
    U = \prod_{k=1}^N T_k(\phi_k)
\end{equation}
During physical execution, the exact phase imparted by each component deviates from the ideal setting. We decompose this microscopic phase error into a time-independent systematic calibration offset and a zero-mean dynamic temporal fluctuation:
\begin{equation}
    \delta\phi_k = \delta\phi_k^{\text{static}} + \delta\phi_k^{\text{noise}}
\end{equation}
We model the dynamic temporal noise as independent Gaussian random variables with non-uniform variances, $\delta\phi_k^{\text{noise}} \sim \mathcal{N}(0, \sigma_k^2)$, characterizing heterogeneous physical noise sources such as thermal variance or electronic control jitter. The physical transformation realized by the hardware in a single operational snapshot is therefore the perturbed product:
\begin{equation}
    \tilde{U} = \prod_{k=1}^N T_k(\phi_k + \delta\phi_k^{\text{static}} + \delta\phi_k^{\text{noise}})
\end{equation}

By isolating the target phase from the errors, we can express each local transformation using its Hermitian phase generator $G_k$. To factor out the ideal target unitary $U$, we commute the error generators to the right side of the ordered product. This commutation dresses the local generators with the subsequent ideal phase transformations, yielding effective configuration-dependent error generators:
\begin{equation}
    \tilde{G}_k(\boldsymbol{\phi}) = \left( \prod_{j=k}^N T_j(\phi_j) \right)^\dagger G_k \left( \prod_{j=k}^N T_j(\phi_j) \right)
\end{equation}

The complete physical unitary matrix separates into the ideal transformation and an exponential containing both static and dynamic error terms. Ignoring higher-order commutators between the small error generators, we decouple the static distortion from the dynamic noise:
\begin{equation}
    \tilde{U} \approx U \exp\left(i \sum_{k=1}^N \delta\phi_k^{\text{static}} \tilde{G}_k(\boldsymbol{\phi})\right) \exp\left(i \sum_{k=1}^N \delta\phi_k^{\text{noise}} \tilde{G}_k(\boldsymbol{\phi})\right)
\end{equation}

We define the configuration-dependent static distortion as $U_c(\boldsymbol{\phi}) = \exp(i \sum_k \delta\phi_k^{\text{static}} \tilde{G}_k(\boldsymbol{\phi}))$ and the dynamic noise operator as $V(\boldsymbol{\phi}) = \exp(i \sum_k \delta\phi_k^{\text{noise}} \tilde{G}_k(\boldsymbol{\phi}))$. For a given target phase configuration $\boldsymbol{\phi}$, the exact probability of detecting a photon in the $i$-th output mode is dictated by the Born rule:
\begin{equation}
    P(b_i=1 | \boldsymbol{\phi}) = \left( U U_c(\boldsymbol{\phi}) V(\boldsymbol{\phi}) \rho V^\dagger(\boldsymbol{\phi}) U_c^\dagger(\boldsymbol{\phi}) U^\dagger \right)_{ii}
\end{equation}

The first statistical step requires evaluating the expectation value over the dynamic temporal noise occurring during the finite measurement window. Expanding the noise operator $V(\boldsymbol{\phi})$ to second order and applying the expectation operator $\mathbb{E}_{\text{noise}}$ over the Gaussian distributions isolates the surviving terms. The linear terms vanish identically due to the zero-mean nature of the fluctuations. The independence of the spatial noise sources removes all cross-terms, satisfying $\mathbb{E}[\delta\phi_k^{\text{noise}} \delta\phi_j^{\text{noise}}] = \sigma_k^2 \delta_{kj}$. This reduces the noise interaction to a configuration-dependent dephasing map $\mathcal{N}_{\boldsymbol{\phi}}(\rho)$ acting on the state:
\begin{equation}
    \mathcal{N}_{\boldsymbol{\phi}}(\rho) \approx \rho - \frac{1}{2} \sum_{k=1}^N \sigma_k^2 [\tilde{G}_k(\boldsymbol{\phi}), [\tilde{G}_k(\boldsymbol{\phi}), \rho]]
\end{equation}

The second statistical step encompasses the classical shadow protocol, which necessitates averaging over an extensive ensemble of random unitary operations drawn from the Haar measure. Uniformly sampling the target matrix $U$ from the $\mathcal{U}(d)$ group mandates that the required phase settings $\boldsymbol{\phi}$ continuously scan the entire parameter space. Averaging the density matrix over this ensemble defines the global effective quantum channel:
\begin{equation}
    \Phi(\rho) = \int d\boldsymbol{\phi} P(\boldsymbol{\phi}) \mathcal{N}_{\boldsymbol{\phi}}(\rho)
\end{equation}

Because the Haar measure is left- and right-invariant, the ensemble of intermediate matrices within $\tilde{G}_k(\boldsymbol{\phi})$ rotates each local hardware generator isotropically across the entire Hilbert space. Integrating this multi-mode dephasing double-commutator over a unitarily invariant distribution constitutes a fundamental twirling operation. According to Schur's Lemma, twirling projects any quantum channel onto the commutant of the unitary group, inevitably contracting the state toward the maximally mixed state. Consequently, the integration of each individual $k$-th term yields a discrete depolarizing channel with a severity proportional to its specific physical variance $\sigma_k^2$. Because the sum of depolarizing channels is itself a depolarizing channel, the spatially heterogeneous noise collapses entirely into a single, homogeneous global expression:
\begin{equation}
    \Phi(\rho) = (1-p)\rho + \frac{p}{d}\mathbb{I}
\end{equation}

The macroscopic depolarization parameter $p$ cleanly absorbs the microscopic Gaussian variances $\sigma_k^2$, totally removing the error model's dependence on any specific spatial index $k$. Simultaneously, the Haar average projects the configuration-dependent static distortion $U_c(\boldsymbol{\phi})$ onto its dominant, non-vanishing mean component $U_c = \mathbb{E}_{\text{Haar}}[U_c(\boldsymbol{\phi})]$. This extracted matrix represents the persistent, configuration-independent coherent rotational bias of the hardware.

Substituting the macroscopic depolarizing channel and the ensemble-averaged static distortion into the probability expression evaluates the combined measurement statistics. Because the maximally mixed term is invariant under any subsequent unitary rotation, the application of $U U_c$ leaves it unaffected:
\begin{equation}
    P(b_i=1) = \left( U U_c \left[ (1-p)\rho + \frac{p}{d}\mathbb{I} \right] U_c^\dagger U^\dagger \right)_{ii}
\end{equation}

Expanding the linear terms within the diagonal element immediately yields the final expected distribution utilized in our phenomenological reconstruction framework:
\begin{equation}
    P(b_i=1) = (1-p) \left( U U_c \rho U_c^\dagger U^\dagger \right)_{ii} + \frac{p}{d}
\end{equation}

\bibliography{bibliography}

\end{document}